\documentclass[12pt]{article}
\pdfoutput=1
\usepackage[DIV13]{typearea}
\usepackage{indentfirst}
\usepackage{amsmath}
\usepackage{mathrsfs}
\usepackage{amssymb}
\usepackage{bbm}
\usepackage{graphicx}
\usepackage{slashed}
\usepackage{multicol}
\usepackage[usenames,dvipsnames]{color}
\usepackage{cite}
\usepackage{cancel}
\usepackage[normalem]{ulem}
\usepackage{array}
\usepackage{multirow}
\RequirePackage[colorlinks=true,urlcolor=blue,anchorcolor=blue,citecolor=blue,filecolor=blue,
linkcolor=blue,menucolor=blue,linktocpage=true,pdfproducer=medialab]{hyperref}
\newcommand{\email}[1]{\href{mailto:#1}{\tt #1}}

\numberwithin{equation}{section}
\newcommand{\LL}{\mathscr{L}}

\def\cD{{\cal D}}
\def\cF{{\cal F}}
\def\cG{{\cal G}}
\def\cH{{\cal H}}
\def\cK{{\cal K}}

\def\cO{{\cal O}}

\def\cP{{\cal P}}
\def\cR{{\cal R}}
\def\cX{{\cal X}}

\def\cL{{\cal L}}

\def\Tr{{\rm Tr}}

\def\iR{{\tt R}}
\def\iL{{\tt L}}
\def\iB{{\tt B}}
\def\be{\begin{equation}}
\def\ee{\end{equation}}
\def\beq{\begin{equation}}
\def\eeq{\end{equation}}
\def\bc{\begin{center}}
\def\ec{\end{center}}
\def\bea{\begin{eqnarray}}
\def\eea{\end{eqnarray}}

\def\nn{\nonumber}
\newcommand{\TeV}{\;\text{TeV}}

\newcommand{\mean}[1]{\langle#1\rangle}
\newcommand{\derp}{\partial}

\newcommand{\unity}{\mathbbm{1}}
\newcommand{\UH}{\mathbf{U}}
\newcommand{\SH}{\mathbf{\Sigma}}

\newcommand{\TL}{\mathbf{T}}
\newcommand{\VL}{\mathbf{V}}
\newcommand{\DL}{\mathbf{D}}
\newcommand{\WL}{\mathbf{W}}
\newcommand{\BL}{\mathbf{B}}

\newcommand{\pH}{\boldsymbol\pi}
\newcommand{\gtt}{\mathfrak{g}}
\newcommand{\htt}{\mathfrak{h}}

\newcommand{\tr}{\Tr}
\newcommand{\raw}{\rightarrow}

\newcommand{\cAt}{\widetilde{{\cal A}}}
\newcommand{\ct}{\tilde{c}}

\newcommand{\VLt}{\widetilde{\mathbf{V}}}
\newcommand{\BLt}{\widetilde{\mathbf{B}}}
\newcommand{\WLt}{\widetilde{\mathbf{W}}}
\newcommand{\SLt}{\widetilde{\mathbf{S}}}

\newcommand{\vh}{\langle \varphi \rangle}
\newcommand{\alf}{\left[\dfrac{\varphi}{f}\right]}
\newcommand{\alfm}{\left[\dfrac{\varphi}{2f}\right]}
\newcommand{\alfd}{\left[\dfrac{2\varphi}{f}\right]}

\newcommand{\blue}[1]{\color{blue} #1 \color{black} }

\begin{document}
\begin{titlepage}
\vspace*{-1cm}
\phantom{hep-ph/***} 
{\flushleft
{\blue{FTUAM-14-30}}
\hfill{\blue{IFT-UAM/CSIC-14-072}}
\hfill{\blue{DFPD-2014/TH/16}}}

\vskip 1.5cm
\begin{center}
{\LARGE\bf Sigma Decomposition}\\[3mm]
\vskip .3cm
\end{center}
\vskip 0.5  cm
\begin{center}
{\large R.~Alonso}~$^{a)}$,
{\large I.~Brivio}~$^{b)}$
{\large M.B.~Gavela}~$^{b)}$,
{\large L.~Merlo}~$^{b)}$,
{\large and S.~Rigolin}~$^{c)}$
\\
\vskip .7cm
{\footnotesize
$^{a)}$~
Department of Physics, University of California at San Diego,\\
 9500 Gilman Drive, La Jolla, CA 92093-0319, USA\\
\vskip .1cm
$^{b)}$~
Departamento de F\'isica Te\'orica and Instituto de F\'{\i}sica Te\'orica, IFT-UAM/CSIC,\\
Universidad Aut\'onoma de Madrid, Cantoblanco, 28049, Madrid, Spain\\
\vskip .1cm
$^{d)}$~
Dipartimento di Fisica ``G.~Galilei'', Universit\`a di Padova and \\
INFN, Sezione di Padova, Via Marzolo~8, I-35131 Padua, Italy
\vskip .5cm
\begin{minipage}[l]{.9\textwidth}
\begin{center} 
\textit{E-mail:} 
\email{ralonsod@ucsd.edu},
\email{ilaria.brivio@uam.es},
\email{belen.gavela@uam.es},
\email{luca.merlo@uam.es},
\email{stefano.rigolin@pd.infn.it}.
\end{center}
\end{minipage}
}
\end{center}

\vskip 1cm
\abstract{ 
In composite Higgs models the Higgs is a pseudo-Goldstone boson of a high-energy strong dynamics.
We have constructed the effective chiral Lagrangian for a generic symmetric coset, restricting to CP-even bosonic operators up to four momenta which turn out to depend on seven parameters, aside from kinetic terms. Once the same sources of custodial symmetry breaking as in the Standard Model are considered, the total number of operators in the basis increases up to ten, again aside from kinetic terms. Under these assumptions, we have then particularised the discussion to three distinct frameworks: the original $SU(5)/SO(5)$ Georgi-Kaplan model, the minimal custodial-preserving $SO(5)/SO(4)$ model and the minimal $SU(3)/(SU(2)\times U(1))$ model, which intrinsically breaks custodial symmetry. The projection of the high-energy electroweak effective theory into the bosonic sector of the Standard Model is shown to match the low-energy chiral effective Lagrangian for a dynamical Higgs, and it uncovers strong relations between the operator coefficients. Finally, the relation with the bosonic basis of operators describing  linear realisations of electroweak symmetry breaking is clarified.}
\end{titlepage}
\setcounter{footnote}{0}

\tableofcontents

%
%
\newpage
\section{Introduction}
A new resonance with mass around 125 GeV has been firmly established at LHC~\cite{Aad:2012tfa,Chatrchyan:2012ufa}. Current data do not indicate deviations from the hypothesis of the Standard Model (SM) Higgs particle \cite{Englert:1964et,Higgs:1964ia,Higgs:1964pj} being a component of the $SU(2)_L$ scalar boson doublet of the electroweak (EW) gauge symmetry. Moreover, even after the LHC 14 TeV upgrade, in the absence of exotic resonances it will not be possible to convincingly establish neither the nature of electroweak symmetry breaking (EWSB), nor the mechanism that protects the Higgs mass from large quadratic corrections.

To stabilise the Higgs mass and cure the electroweak hierarchy problem, two main frameworks for beyond the Standard Model (BSM) physics are commonly considered and still viable with the present data: the underlying high-energy dynamics could be either weakly or strongly interacting. In the first case, the EWSB mechanism is linearly realised, as in the SM, with the physical Higgs being an elementary particle. Even if this possibility is more familiar, the existence of an elementary scalar state would represent a surprising exception, as all known examples of scalar states in nature are understood as being composite. This is indeed the philosophy of the second scenario, in which the EWSB is non-linearly realised and the Higgs arises as a composite particle from the strong dynamics sector. 

The idea of a light composite Higgs originating in the context of a strongly interacting dynamics was first 
developed in the 1980s~\cite{Kaplan:1983fs,Kaplan:1983sm,Banks:1984gj,Georgi:1984ef,Georgi:1984af,Dugan:1984hq} and underwent a recent revival of interest either in strong interacting 4D models~\cite{ArkaniHamed:2001nc,ArkaniHamed:2002qy} or in 5D Ads/CFT versions~\cite{Contino:2003ve,Agashe:2004rs,Contino:2006qr,Gripaios:2009pe}. In this framework, a global symmetry group $\cG$ is postulated at high energies and broken spontaneously by some strong dynamics mechanism to a subgroup $\cH$ at a scale $\Lambda_s$. The characteristic scale of the corresponding Goldstone boson (GB) sector is usually denoted by $f$ and satisfies the relation $\Lambda_s \le 4\pi f$ \cite{Manohar:1983md}. Among this set of GBs, three are usually identified with the would-be-longitudinal components of the SM gauge bosons and one with the Higgs field, $\varphi$. Subsequently, a scalar potential for the Higgs field is dynamically generated, inducing EWSB and providing a (light) mass to the Higgs particle. 
Being the Higgs a pseudo-GB arising from the global symmetry breaking, its mass is protected against quantum corrections of the high-energy symmetric theory, providing an elegant solution to the EW hierarchy problem (see for example Ref.~\cite{Contino:2010rs} for a recent review). The EWSB scale, identified with the vacuum expectation value (vev) of the Higgs field $\mean{\varphi}$ does not need to coincide with the EW scale $v$ defined by the EW gauge boson masses, i.e. the W mass $m_W=g\,v/2$. On the other side, $\mean{\varphi}$ is typically predicted in any specific composite Higgs (CH) model to obey a constraint linking it to the EW scale $v$ and to the GB scale $f$. A model-dependent coefficient for the ratio between the strong dynamics scale and the EW sector scale is usually introduced,
\beq
\xi\equiv (v/f)^2\,,
\label{xi}
\eeq
and it quantifies the degree of non-linearity of the theory.  If the Higgs particle is embedded as an EW doublet in a representation of the high-energy group, in the limit $\xi \ll 1$ the construction converges to the corresponding linear realisations of EWSB for most of the operators, except for few
structures connected to the Goldstone boson nature of $h$.

A general feature of these CH scenarios is the presence of exotic resonances lighter than about $1.5\TeV$ that mainly interact with the third family of quarks~\cite{Matsedonskyi:2012ym,Pomarol:2012qf,Redi:2012ha,Marzocca:2012zn,Panico:2012uw,Pappadopulo:2013vca,Contino:2013gna,Matsedonskyi:2014lla}. At present, however, direct searches at colliders of these states are inconclusive. On the other side, indirect studies are viable strategies to shed some light on BSM physics. Low-energy effects of new physics (NP) can be described in a model-independent way via an effective field theory approach, with effective operators written in terms of SM fields and invariant under the SM symmetries. When considering non-linearly realised EWSB and restricting only to the pure gauge sector, i.e. decoupling the Higgs particle 
from the theory, the most general effective Lagrangian, describing gauge and GB interactions and with an expansion up to four momenta, is the so-called Appelquist-Longhitano-Feruglio (ALF) basis, introduced in Refs.~\cite{Appelquist:1980vg,Longhitano:1980iz,Longhitano:1980tm,Feruglio:1992wf,Appelquist:1993ka}. The first attempts of embedding a light Higgs particle in this context have been proposed in Refs.~\cite{Grinstein:2007iv,Contino:2010mh,Azatov:2012bz}. Subsequently, the complete basis of pure-gauge and gauge-Higgs interactions, that extends the ALF basis including a light Higgs particle, has been presented in Refs.~\cite{Alonso:2012px,Gavela:2014vra}.\footnote{The fermion sector has been discussed at different levels and with different aims in Refs.~\cite{Buchalla:2012qq,Alonso:2012jc,Alonso:2012pz,Buchalla:2013rka}. Moreover, Ref.~\cite{Buchalla:2013rka} contains some inferred criticisms to the results presented in Ref.~\cite{Alonso:2012px}, pointing to some allegedly missing and redundant operators. 
Nevertheless, one of the authors in Ref.~\cite{Buchalla:2013rka} agreed in a private communication on the correctness of Ref.~\cite{Alonso:2012px} under the specific assumptions considered there: the list of operators in Ref.~\cite{Alonso:2012px} is a complete basis when only effects that can be described by pure-gauge and gauge-$h$ chiral operators up to four derivatives are considered.
} 

The effective chiral Lagrangian described in Refs.~\cite{Alonso:2012px,Gavela:2014vra} represents a fundamental tool for Higgs analyses at collider and the related phenomenology has been studied in Refs.~\cite{Brivio:2013pma,Brivio:2014pfa,Gavela:2014vra}, mainly focusing on disentangling a composite Higgs from an elementary one, via the analysis of its couplings. Promising discriminating signals include the decorrelation, in the case of non-linear EWSB, of signals expected to be correlated within a given pattern in the linearly realised one, i.e. between some pure-gauge couplings versus gauge-Higgs ones and also between specific couplings with the same number of external Higgs legs (see also Refs.~\cite{Azatov:2012bz,Isidori:2013cga} for the latter type of decorrelations); furthermore, anomalous signals expected at first order in the non-linear realisation may appear only at higher orders of the linear one, and vice versa. 

In this paper, the focus is on the connection between the high-energy (i.e above the GB scale $f$) effective chiral Lagrangian of specific CH realisations and the low-energy (i.e. below $f$) effective chiral Lagrangian for a dynamical Higgs derived in Ref.~\cite{Alonso:2012px}, restricting to the CP-even bosonic sector. Because of predictivity, an important issue on the analysis of generic symmetric cosets $\cG/\cH$ will be the determination of the number of free parameters in the high-energy theory, which may constrain the freedom of the low-energy one. In particular, three explicit CH realisations will be considered in the following: the original $SU(5)/SO(5)$ Georgi-Kaplan model~\cite{Georgi:1984af}, the minimal intrinsically custodial-preserving $SO(5)/SO(4)$ model~\cite{Agashe:2004rs} and the minimal intrinsically custodial-breaking $SU(3)/(SU(2)\times U(1))$ model. By custodial breaking we mean here sources of breaking other than those resulting from gauging the SM subgroup.

The three models considered exhibit typical features of CH models present in the literature and therefore the results obtained here can be straightforwardly generalised to other contractions. It will be shown that the low-energy effects of any of the considered CH models, irrespective of the chosen symmetric coset ${\cal G}/{\cal H}$, can always be described in terms of effective operators invariant under the SM symmetry and written in terms of SM gauge bosons and a scalar singlet $h$, whose couplings have model-dependent constraints. The existence of peculiar patterns in the coefficients of the low-energy effective chiral operators should indeed provide very valuable information when trying to unveil the nature of the EWSB mechanism. 

The paper has been organised as follows. Sect.~\ref{Sect:ExtendedALF} is devoted to recalling the low-energy effective chiral Lagrangian introduced in Ref.~\cite{Alonso:2012px}. Sect.~\ref{Sect:BasisGeneral} contains the high-energy effective chiral Lagrangian, describing the CP-even interactions among SM gauge bosons and the GBs associated to the symmetric coset $\cG/\cH$. Only operators with at most four derivatives are retained in the Lagrangian. Furthermore, no source of custodial breaking besides the SM ones is considered. In Sects.~\ref{Sect:GKmodel}, \ref{Sect:MinimalSO5} and \ref{Sect:MinimalSU3}, the low-energy effective EW chiral Lagrangian is then derived from the high-energy one for the $SU(5)/SO(5)$, $SO(5)/SO(4)$ and $SU(3)/(SU(2)\times U(1))$ composite Higgs models. Finally in Sect.~\ref{Sect:Matching}, the connection with the EW effective linear Lagrangian is also discussed. Conclusions are presented in Sect.~\ref{Sect:Conc}. Technical details on the construction of the models and on the 
comparison with the literature are deferred to the appendix.

%
%
\boldmath
\section{Electroweak low-energy effective chiral Lagrangian}
\unboldmath
\label{Sect:ExtendedALF}

The scalar sector of the SM is gifted with an accidental $SU(2)_L\times SU(2)_R$ global symmetry spontaneously broken to the diagonal component $SU(2)_C$ after the Higgs field gets a non-vanishing vev. The three $SU(2)_L\times SU(2)_R/SU(2)_C$ GBs, can be described at low-energies by a non-linear $\sigma$-model using a dimensionless unitary field $\UH(x)$. The latter transforms in the bi-doublet representation of the $SU(2)_L\times SU(2)_R$ global group and is defined by
\beq
\UH(x)=e^{i\,\pH(x)/v}\, , \qquad \qquad  \UH(x) \rightarrow L\, \UH(x) R^\dagger\, ,
\label{SMGBs}
\eeq
where $\pH(x)=\pi^a(x)\sigma_a$ (with $\sigma_a$ the usual Pauli matrices) and $L,R$ denoting respectively $SU(2)_{L,R}$ global transformations. Moreover, the covariant derivative of the non-linear field $\UH(x)$ can be written as,
\beq
\DL_\mu \UH(x) \equiv \derp_\mu \UH(x) +ig\WL_{\mu}(x)\UH(x) - \dfrac{ig'}{2} B_\mu(x) \UH(x)\sigma_3 \,,
\eeq
where $\WL_\mu(x)\equiv W_{\mu}^a(x)\sigma_a/2$. From the non-linear field $\UH(x)$ and its covariant derivative $\DL_\mu \UH(x)$, it is possible to define  (pseudo-)scalar and  vector chiral fields transforming in the adjoint of $SU(2)_L$ as follows:
\beq
\begin{aligned}
\TL(x) &\equiv \UH(x) \sigma_3 \UH^\dagger(x)\,,\qquad\qquad &\TL(x)&\rightarrow L\,\TL(x)L^\dagger\,,\\
\VL_\mu(x) &\equiv \left(\DL_\mu \UH(x)\right)\UH^\dagger(x)\,,\qquad\qquad&\VL_\mu(x)&\rightarrow L\,\VL_\mu(x)L^\dagger\,.
\end{aligned}
\label{oldchiral}
\eeq
These two fields $\TL$ and $\VL_\mu$, together with the SM gauge fields $W^a_\mu$ and $B_\mu$ and their derivatives, would suffice as building blocks to construct the EW effective chiral Lagrangian~\cite{Appelquist:1980vg,Longhitano:1980iz,Longhitano:1980tm,Feruglio:1992wf,Appelquist:1993ka}, in the absence of a light Higgs in the low-energy spectrum. Performing an expansion up to four momenta, a complete basis of $SU(2)_L \times U(1)_Y$ invariant CP-even operators -- the ALF basis -- is composed of eighteen independent operators.

The discovery of a light scalar degree of freedom, corresponding to the SM Higgs particle, implies the necessity of extending the ALF basis. The electroweak  chiral effective Lagrangian 
should now describe also other interactions with a CP-even scalar singlet field $h$ that may (or may not) participate in the EWSB mechanism. The extension of the ALF basis  to include a new light scalar degree of freedom in the low-energy chiral Lagrangian (which we will denote by $\cL_{low}$ in what follows), has been derived in Ref.~\cite{Alonso:2012px}, where the complete set of independent CP-even operators describing pure-gauge and gauge-Higgs interactions, up to four derivatives, has been listed\footnote{The complete set of CP-odd operators describing pure-gauge and gauge-Higgs interactions has been presented in Ref.~\cite{Gavela:2014vra}. Chiral interactions including fermions have been considered in \cite{Cvetic:1988ey,Alonso:2012jc,Alonso:2012pz,Buchalla:2013rka}.}. For definiteness and later comparison, we report here the full set of operators, organised by their number of derivatives and their custodial character\footnote{The set of pure-gauge and gauge-Higgs operators in Eqs.~(\ref{ExtALFOp2}) 
and (\ref{ExtALFOp4}) 
exactly matches that in Ref.~\cite{Alonso:2012px}; nevertheless, the labelling of some operators here is different with respect to that in Ref.~\cite{Alonso:2012px} and  matches that in Ref.~\cite{Brivio:2013pma} instead.}:\begin{description}
\item[Operators with two derivatives]
\beq
\hspace{-1cm}
\begin{aligned}
&\underline{\text{Custodial preserving}}\hspace{1.8cm}
&&\underline{\text{Custodial breaking}}\\
\cP_C &= -\frac{v^2}{4}\tr(\VL^\mu \VL_\mu) \hspace{1.8cm}
&\cP_T &= \frac{v^2}{4} \tr(\TL\VL_\mu)\tr(\TL\VL^\mu)
\end{aligned}
\label{ExtALFOp2}
\eeq
\item[Operators with four derivatives]
\beq
\hspace{-0.5cm}
\begin{aligned}
&\underline{\text{Custodial preserving}}\hspace{1.8cm} &&\underline{\text{Custodial breaking}}\\
\cP_B&=-\dfrac{1}{4} B_{\mu\nu}B^{\mu\nu} 
&\cP_{12} &= g^2 (\tr(\TL\WL_{\mu\nu}))^2\\
\cP_W&=-\dfrac{1}{2}\tr(\WL_{\mu\nu}\WL^{\mu\nu})
&\cP_{13} &= ig \tr(\TL\WL_{\mu\nu})\tr(\TL[\VL^\mu,\VL^\nu]) \\
\cP_{1} &= gg' B_{\mu\nu}\tr( \TL\WL^{\mu\nu}) 
&\cP_{14} &= g \epsilon_{\mu\nu\rho\lambda} \tr(\TL\VL^\mu) \tr(\VL^\nu \WL^{\rho\lambda})\\
\cP_{2} &= ig' B_{\mu\nu}\tr(\TL[\VL^\mu,\VL^\nu]) 
&\cP_{15} &= \tr(\TL\cD_\mu\VL^\mu) \tr(\TL\cD_\nu\VL^\nu) \\
\cP_{3} &= ig  \tr(\WL_{\mu\nu} [\VL^\mu,\VL^\nu]) 
&\cP_{16} &= \tr([\TL,\VL_\nu]\cD_\mu\VL^\mu) \tr(\TL\VL^\nu) \\
\cP_{4} &= ig' B_{\mu\nu}\tr(\TL\VL^\mu) \derp^\nu (h/v)
&\cP_{17} &= ig \tr(\TL\WL_{\mu\nu}) \tr(\TL\VL^\mu) \derp^\nu (h/v) \\
\cP_{5} &= ig  \tr(\WL_{\mu\nu}\VL^\mu) \derp^\nu (h/v)
&\cP_{18} &= \tr(\TL[\VL_\mu,\VL_\nu])\tr(\TL\VL^\mu) \derp^\nu(h/v) \\
\cP_{6} &= (\tr(\VL_\mu\VL^\mu))^2 
&\cP_{19} &= \tr(\TL\cD_\mu\VL^\mu)\tr(\TL\VL_\nu) \derp^\nu(h/v)   \\
\cP_{7} &= \tr(\VL_\mu\VL^\mu) \derp_\nu\derp^\nu(h/v)  
&\cP_{21} &= (\tr(\TL\VL_\mu))^2 \derp_\nu(h/v)\derp^\nu(h/v)  \\
\cP_{8} &= \tr(\VL_\mu\VL_\nu) \derp^\mu(h/v)\derp^\nu(h/v) \qquad 
&\cP_{22} &= \tr(\TL\VL_\mu)\tr(\TL\VL_\nu) \derp^\mu(h/v)\derp^\nu(h/v) \\
\cP_{9} &= \tr((\cD_\mu\VL^\mu)^2) 
&\cP_{23} &= \tr(\VL_\mu\VL^\mu) (\tr(\TL\VL_\nu))^2\\
\cP_{10} &= \tr(\VL_\nu\cD_\mu\VL^\mu) \derp^\nu(h/v) 
&\cP_{24} &= \tr(\VL_\mu\VL_\nu)\tr(\TL\VL^\mu)\tr(\TL\VL^\nu)\\
\cP_{11} &= (\tr(\VL_\mu\VL_\nu))^2 
&\cP_{25} &= (\tr(\TL\VL_\mu))^2 \derp_\nu\derp^\nu (h/v) \\
\cP_{20} &= \tr(\VL_\mu\VL^\mu) \derp_\nu(h/v)\derp^\nu(h/v)
&\cP_{26} &= (\tr(\TL\VL_\mu)\tr(\TL\VL_\nu))^2\,. 
\end{aligned}
\label{ExtALFOp4}
\eeq
\end{description}
In Eqs.~(\ref{ExtALFOp2}) and (\ref{ExtALFOp4}), the operators have been classified according to their custodial character: those on the right column, indicated as ``custodial breaking'', describe tree-level effects of custodial breaking sources beyond the SM (gauge) ones. All these operators are easily identified by the presence of the scalar chiral field $\TL(x)$ not in association with the $B_{\mu\nu}$ field strength. The ALF basis can simply be obtained from Eqs.~(\ref{ExtALFOp2}) and (\ref{ExtALFOp4}), disregarding all the operators containing derivatives of $h$. In Eq.~(\ref{ExtALFOp4}), $\cD_\mu$ denotes the covariant derivative in the adjoint representation of 
$SU(2)_L$, i.e.
\beq
\cD_\mu\VL_\nu\equiv \derp_\mu \VL_\nu + i\,g\,\left[\WL_\mu,\VL_\nu\right]\,.
\eeq

To fully encompass the $h$ sector, this list should be extended by a set of four pure-$h$ operators: 
\begin{description}
\item[Operators with two derivatives]
\beq
\cP_{H}=\dfrac{1}{2}\left(\derp_\mu h\right)^2\,.
\label{ExtALFOph}
\eeq
\item[Operators with four derivatives]
\beq
\begin{gathered}
\cP_{\square H}=\dfrac{1}{v^2}\left(\derp_\mu\derp^\mu h\right)^2\,,\qquad\qquad
\cP_{\Delta H}=\dfrac{1}{v^3}\left(\derp_\mu h\right)^2\square h\,,\\
\cP_{DH}=\dfrac{1}{v^4}\left((\derp_\mu h)(\derp^\mu h) \right)^2\,.
\end{gathered}
\label{ExtALFOph2}
\eeq
\end{description}

In summary, the low-energy electroweak chiral Lagrangian describing the CP-even gauge-Goldstone and the gauge-scalar interactions can thus be written as
\beq
\LL_\text{low}=\LL^{p^2}_\text{low}+\LL^{p^4}_\text{low}\,,
\label{LALFgen}
\eeq
where  $\LL^{p^2}_\text{low}$ and $\LL^{p^4}_\text{low}$ contain two and and four-derivative operators,
\beq
\begin{aligned}
\LL^{p^2}_\text{low} =&\cP_C\cF_C(h)+c_T\cP_T\cF_T(h)+\cP_H\cF_H(h)\,,
\label{LALF}\\
\LL^{p^4}_\text{low} =&\cP_B\cF_B(h)+\cP_W\cF_W(h)+\sum_{i=1}^{26}c_i\,\cP_i\cF_i(h)+\\
&+c_{\square H}\cP_{\square H}\cF_{\square H}(h)+c_{\Delta H}\cP_{\Delta H}\cF_{\Delta H}(h)+c_{DH}\cP_{DH}\cF_{DH}(h)\,,
\end{aligned}
\eeq
with the functions $\cF_i(h)$ encoding a generic dependence on  $h$ (in particular, no derivatives of $h$ are included in  $\cF_i(h)$). 

The effective Lagrangian in Eq.~(\ref{LALFgen}) describes at low-energy (EW scale $v$) any model with a light CP-even Higgs, focusing only on the bosonic sector and restricting to CP-even operators with at most four derivatives. Indeed,  $\LL_\text{low}$  describes an extended class of ``Higgs'' models, ranging from the SM scenario to  technicolor-like ansatzs and intermediate situations such as dilaton-like scalar frameworks and CH models. 

Notice that here, for later convenience, a slightly different notation is adopted with respect to that in Refs.~\cite{Alonso:2012px,Brivio:2013pma} for the definition of the $\cF_i(h)$ functions. Here the $\cF_i(h)$ functions are not part of the definition of the operators $\cP_i$, but instead are left outside as multiplicative terms in the Lagrangian. The only dependence on $h$ left inside the operators $\cP_i$ is that corresponding to  derivatives of $h$. Furthermore, in Refs.~\cite{Alonso:2012px,Brivio:2013pma} the dependence on the parameter $\xi$ was made explicit at the Lagrangian level in order to show the connection with the linear effective Lagrangian. Here instead, the $\xi$ weights are reabsorbed in the coefficients $c_i$ and in the functions $\cF_i(h)$. The role of $\xi$ will become clear in the following sections, once specific dynamical Higgs models will be considered. 

According to  NDA~\cite{Manohar:1983md,Jenkins:2013sda}, the weight in front of each four-derivative operator is estimated to be $f^2/\Lambda_s^2\gtrsim1/(4\pi)^2$. This is true for all terms above even if obscured for those operators in Eq.~(\ref{ExtALFOp4}) which include $\partial(h/v)$: their associated $c_i$ have already absorbed a dependence on $\xi$, as mentioned above. To illustrate this, let us consider the example of $\cP_5$:  on physics grounds, factors of $h$ are expected to enter the operators weighted down by $f$,  which is the associated Goldstone boson scale, and the ``natural" definition of the operator would have been
\beq
\cP_{5} = ig  \tr(\WL_{\mu\nu}\VL^\mu) \derp^\nu (h/f)\,,
\eeq
for whose coefficient NDA would  indicate a $f^2/\Lambda_s^2$ weight. Now, the operator definition chosen with $ \derp^\nu (h/v)$ instead of  $\derp^\nu (h/f)$, implies that  $c_5$ has already been redefined in order to reabsorb a factor of $\sqrt{\xi}$, and the overall weight expected for the coefficient of the $\cP_5$ operator as defined in Eq.~(\ref{ExtALFOp4}) is $c_5\sim\sqrt{\xi} f^2/\Lambda_s^2\gtrsim (v/f)\times1/(4\pi)^2 $.

%
%
\boldmath
\section{Effective chiral Lagrangian for symmetric cosets}
\unboldmath
\label{Sect:BasisGeneral}

This section is dedicated to the construction of the high-energy effective Lagrangian in a generic CH setup: a global symmetry group $\cG$ is spontaneously broken by some strong dynamics mechanism at the scale $\Lambda_s$, to a subgroup $\cH$, such that the coset $\cG/\cH$ is symmetric; the minimum requisite is that $dim(\cG/\cH) \ge 4$, i.e. at least four GBs arise from the global symmetry breaking, such that three of them would be then identified with the longitudinal components of the SM gauge bosons and one with the light scalar resonance observed at LHC. No fermionic operators will be considered, and only CP-even ones will be retained among the set of bosonic operators, up to four derivatives.

This generic effective chiral Lagrangian for symmetric cosets will be applied in the subsequent sections to specific CH models.

\boldmath
\subsection{Non-linear realisations of the $\cG/\cH$ symmetry breaking}
\unboldmath

Following the general CCWZ construction \cite{Coleman:1969sm,Callan:1969sn}, the GB degrees of freedom arising from the global symmetry breaking of the group $\cG$ down to the subgroup $\cH$ can be described by the field $\Omega(x)$:
\beq
\Omega(x)\equiv e^{i\,\Xi(x)/2f}\,,
\label{DefOmega}
\eeq
transforming under the global groups $\cG$ and $\cH$ as\footnote{Depending on whether the group $SU(N)$ or $SO(N)$ is considered, $\htt^{-1}=\htt^{\dagger}$ or $\htt^{-1}=\htt^{T}$ should be used, respectively.}
\beq
\Omega(x)\rightarrow \gtt\, \Omega(x)\, \htt^{-1}(\Xi, \gtt)\,,
\label{TransOmega1}
\eeq
where $\gtt$ is a (global) element of $\cG$ while $\htt(\Xi, \gtt)$ is a (local) element of $\cH$ depending explicitly on $\gtt$ and on the Goldstone boson field $\Xi(x)$. For the sake of brevity,  in what follows it will be understood $\htt\equiv\htt(\Xi, \gtt)$ unless otherwise stated. Eq.~(\ref{TransOmega1}) defines the non-linear transformation of $\Xi(x)$. Denoting by $T_a$ (with $a=1,\ldots,dim(\cH)$) the generators of $\cH$ and by $X_{\hat{a}}$ (with $\hat{a}=1,\ldots,dim(\cG/\cH)$) the generators of the coset $\cG/\cH$ in such a way that $\left(T_a,X_{\hat{a}}\right)$ form an orthonormal basis of $\cG$, the GB field matrix explicitly reads:
\beq
\Xi(x)=\Xi^{\hat{a}}(x) \,X_{\hat{a}}\,.
\eeq

In all realistic models considered in the literature, either in the context of QCD,  EW chiral Lagrangian or  CH models, the generators satisfy the following schematic conditions: 
\beq
\left[T,\, T \right] \propto T\,,\qquad\qquad
\left[T,\, X \right]\propto X\,,\qquad\qquad
\left[X,\, X \right] \propto T\,,
\label{SymCoset}
\eeq
the last one being the condition for a symmetric coset\footnote{The first condition follows from $\cH$ being closed. The second one can be deduced from the first one together with the fact that for compact groups the structure constants are completely antisymmetric.}.
In other words, a symmetric $\cG/\cH$ coset admits the automorphism (usually dubbed ``grading'') $\gtt\raw \cR(\gtt)=\gtt_\cR$,
\beq
\cR:\quad
\begin{cases}
T_a\rightarrow +T_a\\
X_{\hat{a}}\rightarrow -X_{\hat{a}}
\end{cases}
\eeq
consistent with the commutation relations in Eq.~(\ref{SymCoset}). For instance, in the case of chiral groups like $SU(N)_L\times SU(N)_R \raw SU(N)_V$ this grading corresponds to the parity operator that leaves invariant the vector generators, while changing the sign to the axial-vector ones. 

As already pointed out in Ref.~\cite{Coleman:1969sm}, it can be shown that in the presence of such an automorphism the non-linear field transformations of $\Omega(x)$ can also be recast as:
\beq
\Omega(x)\rightarrow \htt\, \Omega(x)\, \gtt_{\cR}^{-1}\,.
\label{TransOmega2}
\eeq
From Eqs.~(\ref{TransOmega1}) and (\ref{TransOmega2}), it is thus possible to define for all symmetric cosets a ``squared'' non-linear field $\SH(x)$:
\beq
\SH(x)\equiv \Omega(x)^2\,,
\label{Sigma} 
\eeq
transforming under $\cG$ as,
\beq
\SH(x)\rightarrow \gtt\, \SH(x)\, \gtt_{\cR}^{-1}\,,
\eeq
showing explicitly that the transformation on $\Xi(x)$ is a realisation of $\cG$, and that it is linear when restricted to $\cH$. Notice that the GB field matrix $\SH(x)$ transforms under the grading $\cR$ as:
\beq
\SH(x) \rightarrow \SH(x)^{-1}\,.
\eeq

It is then a matter of taste, in a symmetric coset framework, to use $\Omega(x)$ or $\SH(x)$ for describing the GBs degrees of freedom and the interactions between the GB fields and the gauge/matter fields. The $\Omega$-representation to derive $\cH$-covariant quantities entering the model Lagrangian has been used in several examples. However, when discussing QCD or EW chiral Lagrangians, the $\Sigma$-representation has been more often adopted. To make a straightforward comparison with $\LL_{\text{low}}$ introduced in Sect.~\ref{Sect:ExtendedALF}, the $\Sigma$-representation will be kept in the following. 

One can introduce the vector chiral field\footnote{In order to avoid confusion we will denote with ``$\sim$'' gauge bosons and chiral fields embedded in $\cG$.}:
\beq
\VLt_\mu = \left(\partial_\mu \SH \right) \SH^{-1}\,, \qquad\qquad 
\VLt_\mu \rightarrow \gtt \, \VLt_\mu \,\gtt^{-1}\,,
\label{newchiral}
\eeq
transforming in the adjoint of $\cG$. The effective Lagrangian describing the GB interactions in the context of the non-linearly realised $\cG$ breaking mechanism, with symmetric coset $\cG/\cH$, can then be constructed solely from $\VLt_\mu$. 

In a realistic context, however, gauge interactions should be introduced, and to assign quantum numbers it is convenient to formally gauge the full group $\cG$. In the symmetric coset case, it is possible to define both the $\cG$ gauge fields $\SLt_\mu$, and the graded siblings $\SLt_{\mu}^{\cal R} \equiv \cR (\SLt_{\mu})$, transforming under $\cG$, respectively, as:
\beq
\SLt_{\mu} \rightarrow \gtt \, \SLt_{\mu} \, \gtt^{-1} - \frac{i}{g_S} \gtt (\partial_\mu \,\gtt^{-1}) \,,
\qquad\qquad
\SLt_{\mu}^{\cal R} \rightarrow  \gtt_\cR \, \SLt_{\mu}^{\cal R}\, \gtt^{-1}_\cR - \frac{i}{g_S} \gtt_\cR (\partial_\mu \,\gtt^{-1}_\cR) \,,
\label{TransS2}
\eeq
with  $g_S$ denoting the associated gauge coupling constant.
The (gauged) version of the chiral vector field $\VLt_\mu$ can then be defined as: 
\beq
\VLt_\mu = \left(\DL_\mu \SH \right) \SH^{-1} \,,
\label{newchiralg}
\eeq
with the covariant derivative of the non-linear field $\SH(x)$ being,
\beq
\DL_\mu\SH =\derp_\mu \SH + i\,g_S (\SLt_\mu \SH - \SH\, \SLt^\cR_\mu)\,.
\eeq
The following three $\cG$-covariant objects can thus be used as building blocks for the (gauged) effective chiral Lagrangian:
\beq
\VLt_\mu  \,, \qquad 
\SLt_{\mu\nu} \quad 
\text{and} \quad 
\SH\, \SLt_{\mu\nu}^\cR \, \SH^{-1}\,.
\eeq
The introduction of the graded vector chiral field $\VLt_\mu^\cR$ does not add any further independent structure, as indeed
\beq
\VLt_\mu^\cR \equiv \cR (\VLt_\mu) = \left(\DL_\mu \SH \right)^{-1} \SH 
\qquad  {\rm with} \qquad 
\SH\,\VLt_\mu^\cR\,\SH^{-1} = -\VLt_\mu \,.
\label{Vgrading}
\eeq

\subsection{Basis of independent operators}
It is now possible to derive the most general operator basis describing the interactions of the $\cG$ gauge fields and of the GBs of a non-linear realisation of the symmetric coset $\cG/\cH$. Performing an expansion in momenta and considering CP even operators with at most four derivatives, one obtains the following nine independent operators: 
\begin{description}
\item[2-momenta operator]
\beq
\Tr\left(\VLt_\mu\VLt^\mu\right)\,.
\label{GbasisV}
\eeq
This operator describes the kinetic terms for the GBs and, once the gauge symmetry is broken, results in masses for those GBs associated to the broken generators.
\item[4-momenta operators with explicit gauge field strength $\SLt_{\mu\nu}$]
\beq
\Tr\left(\SLt_{\mu \nu}\SLt^{\mu \nu}\right)\,,\qquad\qquad
\Tr\left(\SH\,\SLt^R_{\mu\nu}\,\SH^{-1}\,\SLt^{\mu\nu}\right)\,,\qquad\qquad
\Tr\left(\SLt_{\mu\nu}\left[\VLt^\mu,\VLt^\nu\right]\right)\,.
\label{Gbasis_S}
\eeq
The first operator describes the kinetic terms for the gauge bosons $\SLt_\mu$. The other two contain gauge-GB and pure-gauge interactions.
\item[4-momenta operators without explicit gauge field strength $\SLt_{\mu\nu}$]
\beq
\begin{gathered}
\Tr\left(\VLt_\mu\,\VLt^\mu\right)\Tr\left(\VLt_\nu\,\VLt^\nu\right)\,,\qquad\qquad
\Tr\left(\VLt_\mu\,\VLt_\nu\right) \Tr\left(\VLt^\mu\,\VLt^\nu\right)\,, \\
\Tr\left((\cD_\mu\VLt^\mu)^2 \right)\,,\qquad\qquad
\Tr\left(\VLt_\mu\,\VLt^\mu\VLt_\nu\,\VLt^\nu\right) \,,\qquad
\Tr\left(\VLt_\mu\,\VLt_\nu\VLt^\mu\,\VLt^\nu\right)\,,
\end{gathered}
\label{GbasisVV}
\eeq
where the adjoint covariant derivative acting on $\VLt^\mu$ is defined as
\beq
\cD_\mu\VLt^\mu= \derp_\mu \VLt^\mu+i\, g_S\left[\SLt_\mu,\VLt^\mu\right]\,. \nn
\eeq
\end{description}
The operators listed in Eqs.~(\ref{GbasisV})--(\ref{GbasisVV}) represent a complete set of independent structures 
describing the interactions among $\cG$ gauge bosons and the GBs associated to $\cG/\cH$ in the $\Sigma$-representation. Additional Lorentz and $\cG$-invariant structures could be {\it a priori} considered beyond those in the previous list (aside from those that are trivially not independent). Of particular interest are the operator $\Tr((\cD_\mu\VLt^\mu) \VLt_\nu\,\VLt^\nu)$ and operators containing determinants. However, the first one is not invariant under the grading $\cR$ (see Eq.~(\ref{Vgrading})) and therefore cannot be retained in the previous set of independent operators. Moreover, invariants of the second type are redundant once restricting only to couplings with at most four derivatives: the Cayley-Hamilton theorem to reduce determinants in terms of traces is a useful tool to prove it. 

It is worth noticing that in specific $\cG/\cH$ realisations, some of the operators listed may not be independent. For example the operators with traces of four $\VLt^\mu$ appearing in the second line of Eq.~(\ref{GbasisVV}) are redundant in the case $\cG=SU(2)_L\times SU(2)_R$ and $\cH=SU(2)_V$, as they decompose in products of traces of two $\VLt^\mu$. It may not be true in models with larger group $\cG$, as it depends on the specific algebra relations of the generators.

Finally, some caution should be also used when fermions are introduced. In this case all operators containing $\cD_\mu\VLt^\mu$ can be traded, via equations of motion, by operators containing fermions and a careful analysis should be performed to avoid the presence of redundant terms.

\boldmath
\subsection{General EW effective Lagrangian for a symmetric $\cG/\cH$ coset}
\unboldmath

The list of operators in Eqs.~(\ref{GbasisV})--(\ref{GbasisVV}) is valid on general grounds when formally gauging the full group $\cG$. Nevertheless, in most realisations of CH models only the SM gauge group is gauged. Consequently, in the generic gauge field $\SLt_\mu$, only the EW components should be retained. While no new operator structures appear in the sector made out exclusively of  $\VLt_\mu$ fields (see Eqs.~(\ref{GbasisV}) and (\ref{GbasisVV})), all operators where the gauge field strength appears explicitly, such as those in Eq.~(\ref{Gbasis_S}), should be ``doubled'' by substituting $\SLt_\mu$ either with $\WLt_\mu$ or $\BLt_\mu$, defined by
\beq
\WLt_\mu\equiv W^a_\mu \,Q^a_L \qquad {\rm and} \qquad \BLt_\mu \equiv B_\mu \,Q_Y\,,
\label{WBSU(5)}
\eeq
where $Q^a_L$ and $Q_Y$  denote the embedding in $\cG$ of the $SU(2)_L\times U(1)_Y$ generators. It follows that a larger number of invariants can be written in this case. In consequence, the CP-even 
 EW high-energy chiral Lagrangian describing up to four-derivative bosonic interactions, $\LL_{\text{high}}$, contains in total thirteen operators:
\beq
\LL_{\text{high}}=\LL^{p^2}_{\text{high}}+\LL^{p^4}_{\text{high}}\,,
\label{LLG}
\eeq
where
\begin{align}
\LL^{p^2}_{\text{high}}&=\cAt_C\,,
\label{LGn2}\\
\LL^{p^4}_{\text{high}}&=\cAt_B+\cAt_W+\ct_{B\Sigma}\cAt_{B\Sigma}+\ct_{W\Sigma}\cAt_{W\Sigma}+\sum_{i=1}^8\ct_i\,\cAt_i\,,
\label{LGn4}
\end{align}
with
\beq
\begin{aligned}
\cAt_C    &= -\frac{f^2}{4}\Tr\left(\VLt_\mu\VLt^\mu\right)\,,\qquad\qquad\qquad
&\cAt_{3} &=i\,g\,\Tr\left(\WLt_{\mu\nu}\left[\VLt^\mu,\VLt^\nu\right]\right)\,,\\
\cAt_B	  &=-\frac{1}{4} \Tr\left(\BLt_{\mu \nu}\BLt^{\mu \nu}\right)\,,
&\cAt_{4} &= \Tr\left(\VLt_\mu\,\VLt^\mu\right)\Tr\left(\VLt_\mu\,\VLt^\mu\right)\,,\\
\cAt_W	  &=-\frac{1}{4} \Tr\left(\WLt_{\mu \nu}\WLt^{\mu \nu}\right)\,,
&\cAt_{5} &= \Tr\left(\VLt_\mu\,\VLt_\nu\right) \Tr\left(\VLt^\mu\,\VLt^\nu\right)\,,\\
\cAt_{B\Sigma} &=g^{\prime 2}\Tr\left(\SH\BLt_{\mu \nu}\SH^{-1}\BLt^{\mu \nu}\right)\,,
&\cAt_{6} &= \Tr\left((\cD_\mu\VLt^\mu)^2 \right)\,,\\
\cAt_{W\Sigma} &=g^2\Tr\left(\SH\WLt_{\mu \nu}\SH^{-1}\WLt^{\mu \nu}\right)\,,
&\cAt_{7} &= \Tr\left(\VLt_\mu\,\VLt^\mu\VLt_\nu\,\VLt^\nu\right)\,,\\
\cAt_{1}  &= g\,g'\,\Tr\left(\SH\BLt_{\mu \nu}\SH^{-1}\WLt^{\mu \nu}\right)\,,
&\cAt_{8} &= \Tr\left(\VLt_\mu\,\VLt_\nu\VLt^\mu\,\VLt^\nu\right)\,,\\
\cAt_{2}  &= i\,g'\,\Tr\left(\BLt_{\mu \nu}\left[\VLt^\mu,\VLt^\nu\right]\right)\,,
&
\end{aligned}
\label{AppelquistBasisG}
\eeq
with the EW covariant derivative  in Eq.~(\ref{AppelquistBasisG})  defined as
\beq
\cD_\mu\VLt^\mu= \derp_\mu \VLt^\mu+i\, g\left[\WLt_\mu,\VLt^\mu\right]+i\, g'\left[\BLt_\mu,\VLt^\mu\right]\,.
\eeq
The coefficients $\ct_i$ are expected to be all of the same order of magnitude, according to the effective field theory approach\footnote{The coefficients of the operators $\cAt_C$, $\cAt_B$ and $\cAt_W$ are taken equal to $1$, which leads to canonical kinetic terms.}.  NDA~\cite{Manohar:1983md,Jenkins:2013sda} applies and indicates that the four-derivative operator coefficients are expected to be of order $f^2/\Lambda_s^2\gtrsim1/(4\pi)^2$.

It is remarkable that, aside from kinetic terms, $\LL_{\text{high}}$ contains only ten independent operators, and thus  at most  ten arbitrary coefficients $\tilde{c_i}$ need to be determined. They will govern the projection of $\LL_{\text{high}}$ into $\LL_{\text{low}}$ (in addition to the parameter(s) of the explicit breaking of the global symmetry).
  
It is also worth to note that the gauging of the SM symmetry breaks explicitly the custodial and the grading symmetries. As a result, custodial and/or grading symmetry breaking operators can arise once quantum corrections induced by SM interactions are considered. 

In the case $\cG=SU(2)_L\times SU(2)_R$ and $\cH=SU(2)_V$, the Lagrangian $\LL_\text{high}$ reduces to the custodial preserving sector of the ALF basis, with the three GBs described by the non-linear realisation of the EWSB mechanism corresponding to the longitudinal degrees of freedom of the SM gauge bosons. In this case, $dim(\cG/\cH) = 3$ and the $h$ field cannot arise as a GB of the spontaneous $\cG$ symmetry breaking.

CH models are, instead, built upon cosets with $dim(\cG/\cH) \ge 4$, the minimal ones being for example 
$SO(5)/SO(4)$ and $SU(3)/(SU(2)\times U(1))$ for the intrinsically custodial preserving and custodial breaking setups, respectively. The four GBs resulting from the non-linear symmetry breaking mechanism  will then correspond to  the three would-be SM GBs and the Higgs particle. In non-minimal models, such as the $SU(5)/SO(5)$ Georgi-Kaplan model, additional GBs appear in the symmetry breaking sector. Either they are light degrees of freedom and then provide interesting candidates for dark matter (see for instance Ref.~\cite{Frigerio:2012uc,Marzocca:2014msa}) or for other  exotic particles, or they should become heavy enough through some ``ad hoc'' global symmetry breaking effect associated to the strong interacting sector~\cite{Georgi:1984af}, leaving a negligible impact on  low-energy physics.  

In the next sections, $\LL_{\text{high}}$ will be particularised to the case of three well-known CH models, by decomposing the field matrix $\SH$ in Eq.~(\ref{Sigma}) into its SM and BSM fields, and projecting into the former: the title of this paper refers to this procedure.

%
%
\boldmath
\section[The \boldmath $SU(5)/SO(5)$ composite Higgs model]{The $SU(5)/SO(5)$ composite Higgs model}
\unboldmath
\label{Sect:GKmodel}

The first CH model was proposed by Georgi and Kaplan~\cite{Georgi:1984af} more than 30 years ago. They assumed   a global $\cG=SU(5)$ symmetry spontaneously broken to the $\cH=SO(5)$ subgroup. It is clearly a non-minimal model as  fourteen GBs arise from the $SU(5)\to SO(5)$ breaking: three of them are then identified with the GBs of the SM, a fourth one with the physical Higgs, while the remaining ten are potentially light states. In Ref.~\cite{Georgi:1984af},  it was shown that strong dynamical effects can induce large (i.e. ${\cal O}(f)$) masses for these extra degrees of freedom, that therefore can be safely disregarded at low-energies. The discussion of this mechanism is beyond the scope of our paper and can be found in that reference. In what follows  all those ten extra GBs are removed from the spectrum and only the three plus one physical degrees of freedom relevant at low energies are considered.

The global $SU(2)_L\times SU(2)_R$ symmetry can be embedded in the unbroken $SO(5)$ residual group and  therefore the model benefits of an approximate custodial symmetry.
 We will first review  in some detail the original  Georgi-Kaplan construction, as this will be the playground for generic (minimal or non-minimal) CH models.

\boldmath 
\subsection{Spontaneous $SU(5)/SO(5)$ symmetry breaking setup}
\unboldmath

Mimicking the detailed discussion presented in Ref.~\cite{Dugan:1984hq}, the spontaneous (global) symmetry breaking pattern $SU(5)\to SO(5)$ can be associated to a scalar field\footnote{This scalar field could result for example from fermionic condensates.} belonging to the symmetric representation of $\cG$ and acquiring a vev $\Delta_0$. The vev can be taken in all generality to be a real, symmetric and orthogonal $5\times 5$ matrix:
\beq
\Delta_0 = \Delta_0^\dagger = \Delta_0^T = \Delta_0^{-1} \,.
\label{Delta0Prop}
\eeq
A convenient choice, which facilitates the identification of the $SU(2)_L\times U(1)_Y$ quantum numbers in the $SU(5)$ embedding, is given by
\beq
\Delta_0 = 
\begin{pmatrix}
 0           & i \sigma_2 & 0 \\
 -i\sigma_2  & 0          & 0 \\
 0           & 0          & 1 \\
\end{pmatrix}\,.
\eeq
It is then possible to describe the massless excitations around the vacuum with a symmetric field $\Delta(x)$, obtained ``rotating'' the vacuum by means of the GB non-linear field $\Omega(x)$: 
\beq
\Delta(x)= \Omega(x)\, \Delta_0\, \Omega(x)^{T}\,,\qquad\qquad
\Delta(x)\rightarrow \gtt\, \Delta(x)\, \gtt^{T}\,.
\label{DeltadefSU5}
\eeq
The field $\Delta(x)$ describes all  fourteen GBs stemming from the $SU(5)/SO(5)$ breaking. Its transformation properties under $SU(5)$, i.e. in the symmetric representation, follow from the invariance of the vacuum under $SO(5)$.  Using the following relations between the vacuum $\Delta_0$ and the broken and unbroken generators,
\beq
\Delta_0 \, T_a \, \Delta_0 = - T_a^T \,,\qquad \qquad 
\Delta_0 \, X_{\hat{a}}\,\Delta_0 = X_{\hat{a}}^T\,,
\label{RelationsDelta0}
\eeq
and because of the relations in Eq.~(\ref{Delta0Prop}), the excitations around the vacuum can be rewritten in terms of the GB field $\SH(x)$:
\beq
\Delta(x) = \Omega(x)^2 \,\Delta_0 \equiv \SH(x)\,\Delta_0\,.
\label{DeltaAndTheta}
\eeq
The vector chiral field $\VLt_\mu$ is then related to the vacuum excitations, 
\beq
\VLt_\mu(x)\equiv\left(\DL_\mu \SH(x)\right) \SH^\dagger(x)=\left(\DL_\mu \Delta(x)\right) \Delta^*(x) \,,
\label{VmutildeTheta}
\eeq 
from which it follows that the GB kinetic term can be written as:
\beq
\Tr \left((\DL_\mu \Delta) (\DL_\mu \Delta)^*\right) = \Tr \left((\DL_\mu \SH) (\DL_\mu \SH)^\dagger\right) = 
   - \Tr\left( \VLt_\mu \VLt^\mu\right)\,.
\eeq

Considering the fourteen GBs arising from the $SU(5)/SO(5)$ breaking and described by $\Omega(x)$ (or $\SH(x)$), the three would-be SM GBs $\cX(x)$ and the scalar singlet field $\varphi(x)$ can be split from the other d.o.f. denoted collectively by $\cK(x)$, by decomposing $\Omega(x)$ as~\cite{Dugan:1984hq}:
\beq
\Omega(x)=e^{i\frac{\varphi(x) }{2f} \cX(x)} e^{i\frac{\cK(x)}{2f}}\,.
\eeq
Strong dynamics effects may induce a heavy mass term for the GBs described by $\cK(x)$~\cite{Georgi:1984af}. The GB field $\Omega(x)$, and $\SH(x)$,  can then be approximated  at energies below $f$ by:
\beq
\Omega(x) \approx e^{i\frac{\varphi(x)}{2f}\cX(x) } \,, \qquad \qquad 
\SH(x)\approx e^{i\frac{\varphi(x)}{f}\cX(x)}\,.
\label{ThetaSU5}
\eeq
Furthermore, the  explicit breaking of the global high-energy symmetry is assumed to induce a potential for the singlet field $\varphi(x)$, which eventually acquires dynamically a non-vanishing vev,
\beq
\frac{\varphi(x)}{f} \equiv \frac{h(x)+\vh}{f} = \left(\frac{h(x)+\vh}{v}\right) \sqrt{\xi}  \,,
\label{varphiExp}
\eeq
where $h(x)$ refers to the physical Higgs (denoted often simply as $h$ in what follows). 

Denoting by $X$ the broken generator along which the EW symmetry breaking occurs,
\beq
X=\dfrac{1}{2}
\begin{pmatrix}
0  & 0 & e_1 \\
0  & 0 & e_2 \\  
e_1^T & e_2^T & 0  \\
\end{pmatrix}
\qquad \qquad \text{with}\quad 
e_1 = \begin{pmatrix} 1 \\ 0 \\  \end{pmatrix}\,, \,\, 
e_2 = \begin{pmatrix} 0 \\ 1 \\  \end{pmatrix}\,,
\eeq
the $SU(5)$ embedding of the SM GB fields can be parametrised as
\beq
\cX(x) =\sqrt2\left(
\begin{array}{ccc}
\UH  &  &  \\
  & \UH &  \\
  &  & 1 \\
\end{array}
\right)X
\left(
\begin{array}{ccc}
\UH^\dag  &  &  \\
  & \UH^\dag &  \\
  &  & 1 \\
\end{array}\right)
=
\dfrac{1}{\sqrt2} 
\begin{pmatrix}
 0 & 0 & \UH(x) e_1 \\
 0 & 0 & \UH(x) e_2  \\
 (\UH(x) e_1)^\dagger & (\UH(x) e_2)^\dagger & 0  \\
\end{pmatrix}\,,  
\label{SU5higgs}
\eeq
with $\UH(x)$ defined in Eq.~(\ref{SMGBs}). In the unitary gauge, $\cX =\sqrt2 X$. Given the peculiar structure of the matrix $X$, the $\SH$ field can be 
written uniquely in terms of linear and quadratic powers of $\cX$ because  $\cX^3=\cX$:
\beq
\SH\equiv \unity + i\sin\left(\frac{\varphi}{f}\right)\, \cX + \left(\cos\left(\frac{\varphi}{f}\right)-1\right) \cX^2\,.
\label{ThetaSU5bis}
\eeq

The last ingredient needed to fully specify the setup is the embedding of the SM fields in $\cG$. Given the choice of vacuum, the 
$SU(2)_L\times U(1)_Y$ generators can be expressed as
\beq
Q^a_L=\dfrac{1}{2}\left(
\begin{array}{ccc}
 \sigma_a &  &  \\
  & \sigma_a &  \\
  &  & 0 \\
\end{array}
\right)\,,\qquad\qquad
Q_Y=\dfrac{1}{2}\left(
\begin{array}{ccc}
 -\unity_2 &  &  \\
  & \unity_2 &  \\
  &  & 0 \\
\end{array}
\right)\,,
\label{Qa}
\eeq
where in these expressions $\sigma_a$ denote the Pauli matrices and the normalisation of the generators  is $\Tr(Q_a Q_a)=1$.

\subsection{The low-energy effective EW chiral Lagrangian}

One can now substitute the explicit expression for $\SH$, $\VLt_\mu$, $\WLt_\mu$ and $\BLt_\mu$ in the operators of the high-energy basis in Eq.~(\ref{AppelquistBasisG}) and obtain $\LL_{\text{low}}$ for the Georgi-Kaplan model as a function of the SM would-be GBs, the light scalar singlet field $\varphi(x)$ and  the SM gauge fields.

\boldmath
\subsubsection{The two-derivative low-energy projection}
\unboldmath

For $SU(5)/SO(5)$, the low-energy projection  of the custodial preserving two-derivative operator reads\cite{Contino:2011np}
\beq
\cAt_C\equiv-\dfrac{f^2}{4}\Tr(\VLt_\mu\VLt^\mu) = \frac{4}{\xi}\sin^2\left[\dfrac{\varphi}{2f}\right]\cP_C\,+\cP_H \,,
\label{LSU5n2}
\eeq
with $\cP_C$ and $\cP_H$ being the operators in $\LL_{\text{low}}$ defined in Eqs.~(\ref{ExtALFOp2}) and (\ref{ExtALFOph}), respectively. Having assumed the absence of any sources of custodial breaking besides the SM ones, no other two-derivative operators arise in the low-energy effective chiral Lagrangian. 

Besides giving rise to the (correctly normalised) $h$ kinetic term described by $\cP_H$, the operator $\cAt_C$ intervenes also in the definition of the SM gauge boson masses. To provide a consistent definition for the SM $W$ mass $m^2_W \equiv g^2 v^2/4$, it is necessary to impose that 
\beq
\xi \equiv \frac{v^2}{f^2} = 4 \sin^2 \frac{\vh}{2f}\,,  
\label{xidefSU5}
\eeq
providing a strict and model-dependent relation between the EW scale $v$, the vev of the scalar field $\varphi$ and the NP scale $f$. Note that in the $\xi \ll 1$ limit the usual SM result $\vh=v$ is recovered. Using Eq.~(\ref{xidefSU5}), the functional dependence on $\varphi/f$ can be nicely translated in terms of the physical $h$ excitation and the EW scale $v$, and the following expressions will be useful later on:
\beq
\begin{aligned}
\sin \left(\frac{\varphi}{ 2f}\right)&=\sin \left(\mbox{arcsin} \left(\frac{v}{ 2f}\right)+\frac{h}{ 2f}\right)=\frac{v}{ 2f}\cos\left(\frac{h}{ 2f}\right)+\sqrt{1-\frac{v^2}{ 4f^2}}\,\sin\left(\frac{h}{ 2f}\right)\,,\\
\cos \left(\frac{\varphi}{ 2f}\right)&=\cos \left(\mbox{arcsin} \left(\frac{v}{ 2f}\right)+\frac{h}{ 2f}\right)=\sqrt{1-\frac{v^2}{ 4f^2}}\cos\left(\frac{h}{ 2f}\right)-\frac{v}{ 2f}\,\sin\left(\frac{h}{ 2f}\right)\,.
\end{aligned}
\eeq

\boldmath
\subsubsection{The four-derivative low-energy projection}
\unboldmath

The low-energy projection of the four-derivative effective operators of Eq.~(\ref{AppelquistBasisG}) gives:
\beq
\begin{aligned} 
\cAt_B=&\,\cP_B \,, \\
\cAt_W=&\,\cP_W \,, \\
\cAt_{B\Sigma}=&-\,4\,g'^2\cos^2\alfm \cP_B  \,, \\
\cAt_{W\Sigma}=&-\,4\,g^2\cos^2\alfm \cP_W \,,\\
\cAt_1 =&\,\sin^2\alfm\cP_1 \,, \\
\cAt_2 =&\,\sin^2\alfm\cP_2\,+\,\sqrt\xi\sin\alf\cP_4 \,,\\
\cAt_3 =&\,2\sin^2\alfm\cP_3\,-\,2\,\sqrt\xi\sin\alf\cP_5\,,\\
\cAt_4 =&\,4\,\xi^2\cP_{DH}\,+\,16\sin^4\alfm\cP_{6} \,-\,16\,\xi\sin^2\alfm\cP_{20} \,,\\
\cAt_5=&\,4\,\xi^2\cP_{DH} \,-\,16\,\xi\sin^2\alfm\cP_{8} \,+\,16 \sin^4\alfm\cP_{11}\,,\\
\cAt_6=&\,-2\,\xi\,\cP_{\square H} \,-\,\dfrac{1}{2}\sin^2\alf\cP_{6} \,+\, 4\,\xi\cos^2\alfm\cP_{8}\,+\, 4\sin^2\alfm\cP_{9} + \\
          &\, -\,2\sqrt{\xi}\sin\alf\left(\cP_{7}-2\cP_{10}\right) \,, 
\label{LSU5n4}
\end{aligned}
\eeq
while the remaining two high-energy operators are not independent when focusing only on the light GBs remaining at low-energies:
\beq
\cAt_7 =\dfrac{1}{4}\left(\cAt_4+\cAt_5\right)\,,\qquad\qquad
\cAt_8 = \dfrac{1}{2}\cAt_5\,.
\eeq
The fact that $\cAt_7$ and $\cAt_8$ do not give independent contributions as they are linear combinations of other high-energy operators is connected with the peculiar structure of the $\cG/\cH$ breaking and has to be inferred case by case. This specific example is similar to the ALF case, where it can be proven that traces of four $\VL_\mu$ can be expressed as products of traces of two $\VL_\mu$. In resume, $\LL_{\text{low}}$ for the $SU(5)/SO(5)$ scenario considered here depends on only eight independent operators, besides the kinetic terms for gauge bosons and GB fields.

It is useful to explicit the dependence on the $\varphi$ field of the expressions in Eq.~(\ref{LSU5n4}), so as to identify easily the correlations to be expected in experimental signals involving  the same number of Higgs external fields, and compare with  those  involving a different number of Higgs particles. To illustrate it, let us momentarily adopt  a slightly different notation for the following operators in $\LL_{\text{low}}$: 
\beq
\begin{aligned}
\cP_i&\equiv \hat\cP_{i\nu}\,\derp^\nu(h/v)\qquad\qquad\text{for}\quad i=4,5,10\,,\\
\cP_7&\equiv \hat\cP_7\,\derp_\nu\derp^\nu(h/v)\,,\\
\cP_8&\equiv \hat\cP_{8\nu\mu}\,\derp^\mu(h/v)\derp^\nu(h/v)\,.
\end{aligned}
\eeq
 The operators $\cAt_2$, $\cAt_3$ and $\cAt_6$ can then be rewritten as
\beq
\begin{aligned}
\cAt_2=&\, \left(\cP_2+2\,\hat\cP_{4\nu}\,\derp^\nu\right)\,\sin^2\alfm\,, \\
\cAt_3=&\, 2\left(\cP_3-2\,\hat\cP_{5\nu}\,\derp^\nu\right)\,\sin^2\alfm\,,\\
\cAt_6=&\,  -2\,\xi\,\cP_{\square H}-\left(\frac{1}{2} \cP_{6}-4\cP_{9} +4\,\hat\cP_{7}\,\derp_\nu\,\derp^\nu -8\,\hat\cP_{10\nu}\,\derp^\nu\right)\,\sin^2\alfm+\\
    &\,+16\,\hat\cP_{8\mu\nu}\,\derp^\mu\sin\left[\dfrac{\varphi}{2f}\right] \,\derp^\nu\sin\left[\dfrac{\varphi}{2f}\right]\,.
\end{aligned}
\label{LEwithG}
\eeq
This decomposition shows that, for any given number  of $\varphi$ external legs,  the gauge interactions stemming -for instance- from $\cP_2$ and $\cP_4$ in $\cAt_2$ combine with a fixed relative weight, independently of the size of $f$ and of the ratio $\langle\varphi\rangle/f$.  That relative weight is equal to that holding for the same set of gauge interactions within the $d=6$ operators of the linear Lagrangian, as it will be discussed in Sect.~\ref{Sect:Matching}. This correlation is intimately related to the fact that the $\varphi$ field was embedded as a $SU(2)_L$ doublet in the high-energy theory. An  analogous discussion applies to $\cAt_3$ and $\cAt_6$ in Eq.~(\ref{LEwithG}).

%
%

\boldmath
\section[The minimal \boldmath$SO(5)/SO(4)$ composite Higgs model]{The minimal $SO(5)/SO(4)$ composite Higgs model}
\unboldmath
\label{Sect:MinimalSO5}
Most of the recent literature in CH models deals with the minimal $SO(5)/SO(4)$ \cite{Agashe:2004rs} setup. The features that make this model appealing are its custodial symmetry approximate conservation and its minimality in terms of number of GBs that arise from the global symmetry breaking: only four to be associated with the SM would-be GBs and the Higgs field.

\boldmath
\subsection{Spontaneous $SO(5)/SO(4)$ symmetry breaking setup}
\unboldmath

The spontaneous $SO(5)/SO(4)$ symmetry breaking can be obtained giving a vev to a scalar field either in a fundamental or in the symmetric adjoint representation. To resemble most the discussion of the $SU(5)/SO(5)$ model  the latter representation is chosen here. Also for this setup, the vacuum can be taken in all generality to be a real, symmetric and orthogonal $5\times 5$ matrix satisfying  Eq.~(\ref{Delta0Prop}), and a convenient choice is to set
\beq
\Delta_0=\begin{pmatrix}
\unity_4  & 0 \\
0  & -1 \\
\end{pmatrix} \,.
\eeq
As in the previous case, it is then possible to describe the massless excitations around the vacuum with a symmetric field $\Delta(x)$ obtained ``rotating'' the vacuum with the GB non-linear field $\Omega(x)$: Eq.~(\ref{DeltadefSU5}) also holds here, with $\mathfrak g$ being now a transformation of $SO(5)$. $\Delta(x)$ transforms in the adjoint of $SO(5)$, as a consequence of the invariance of the vacuum under $SO(4)$, and describes only four GBs. 

 The relations between the vacuum $\Delta_0$ and the broken and unbroken generators presented in Eq.~(\ref{RelationsDelta0}) are valid also for this model, and because of the relations in Eq.~(\ref{Delta0Prop})  the excitations around the vacuum can be reparametrised in the $\Sigma$-representation as in Eq.~(\ref{DeltaAndTheta}), where now $\Omega(x)$ and $\SH(x)$ are given by
\beq
\Omega(x)=e^{i\frac{\varphi(x) }{2f}\cX(x)} \,, \qquad\qquad 
\SH(x) = e^{i\frac{\varphi(x)}{f}\cX(x)}\,.
\label{OmegaThetaSO5}
\eeq
The $SO(5)/SO(4)$ generators can be written in a compact form as
\beq
\left(X_{\hat a}\right)_{ij}=\frac{i}{\sqrt{2}}\left(\delta_{i 5}\delta_{j \hat a}-\delta_{j 5}\delta_{i \hat a}\right)\,,\qquad\qquad 
\hat a=1,\dots,4\,,
\eeq
and denoting the broken generator along which the EW symmetry breaking occurs as $X_{\hat{4}} $,
\bea
X_{\hat{4}} & = \dfrac{i}{\sqrt{2}} \left(
\begin{array}{ccc}
   0 & 0     & 0 \\
   0 & 0     & -e_2 \\
   0 & e_2^T & 0 \\
\end{array} \right)\,,\
\eea
the GB non-linear field reads
\beq
\cX(x) = -\dfrac{i}{\sqrt{2}}\, \Tr\left(\UH \sigma_{\hat{a}}\right) X_{\hat{a}} \,, \qquad \qquad \hat{a}=1,\dots,4\,,
\eeq 
where $\sigma_{\hat{a}}\equiv\left\{\sigma_1,\sigma_2,\sigma_3,i\unity_2\right\}$ and which reduces to $\cX = \sqrt{2}\, X_{\hat 4}$ in the unitary gauge. Alike to the case of the Georgi-Kaplan model, the field $\SH$ takes the simple form in terms of linear and quadratic powers of $\mathcal X$ shown in Eq.~(\ref{ThetaSU5bis}).
 Finally, with this convention the embedding of the $SU(2)_L\times U(1)_Y$ generators in $SO(5)$ reads 
\beq
\begin{aligned}
Q^1_L&=\dfrac{1}{2}\left(
\begin{array}{ccc}
  & -i \sigma_1 &  \\
  i \sigma_1 &  &  \\
  &  & 0 \\
\end{array} \right)\,,\qquad 
& Q^2_L&=\dfrac{1}{2}\left(
\begin{array}{ccc}
  & i \sigma_3 &  \\
  -i \sigma_3 & &  \\
  &  & 0 \\
\end{array} \right)\,,\\
Q^3_L&=\dfrac{1}{2}\left(
\begin{array}{ccc}
 \sigma_2 &  &  \\
  & \sigma_2 &  \\
  &  & 0 \\
\end{array}\right)\,,
\qquad
&Q_Y&=\dfrac{1}{2}\left(
\begin{array}{ccc}
 \sigma_2 &  &  \\
  & -\sigma_2 &  \\
  &  & 0 \\
\end{array} \right)\,.
\end{aligned}
\eeq

\subsection{The low-energy effective EW chiral Lagrangian}

Having chosen the explicit realisation of the $SO(5)/SO(4)$ symmetry breaking mechanism and the representation of the embedding of the SM group charges into $SO(5)$, the substitution of the explicit expressions for $\SH$, $\VLt_\mu$, $\WLt_\mu$ and $\BLt_\mu$ into the operators of the high-energy basis in Eq.~(\ref{AppelquistBasisG}) produces $\LL_{\text{low}}$
for the minimal $SO(5)/SO(4)$ CH model, as a function of the SM would-be GBs and the light scalar resonance $\varphi$.

The low-energy projection of the $SO(5)/SO(4)$ Lagrangian turns out to be exactly the same as that for the $SU(5)/SO(5)$ model. This result depends on the strict connection between $SO(5)$ and $SU(5)$, as indeed the GB matrix fields of the two theories are linked by a unitary global transformation, once decoupling the extra GBs arising in the $SU(5)\to SO(5)$ breaking. Moreover, the gauging of the SM symmetry represents an explicit breaking of the global symmetries and it produces the effect of washing out the differences between the two preserved subgroups, once focusing only on the SM particle spectrum. This also suggests that any model with the minimal number of GBs that can be arranged in a doublet of $SU(2)_L$ and approximate custodial symmetry will yield the same low-energy effective chiral Lagrangian regardless of the specific ultraviolet completion. 

%
%
\boldmath
\section[The \boldmath$SU(3)/(SU(2)\times U(1))$ composite Higgs model]{The $SU(3)/(SU(2)\times U(1))$ composite Higgs model}
\unboldmath
\label{Sect:MinimalSU3}

As a final example, the $SU(3)/(SU(2)\times U(1))$ CH model is now considered. As only four GBs arise from the breaking of the global symmetry, also this model is minimal. However, contrary to the previously discussed CH models, the preserved subgroup $\cH$ does not contain the custodial $SO(4)$ term and therefore no (approximate) custodial symmetry is embeddable in this model. This feature  disfavours phenomenologically the $SU(3)/(SU(2)\times U(1))$ CH model as large tree-level contributions to the $T$ parameter occur. Nevertheless, the study of its low-energy projection is instructive in order to discuss the custodial breaking operators of the effective Lagrangian $\cL_{\text{low}}$ in Eq.~(\ref{LALFgen}). Indeed, although in the initial high-energy $SU(3)/(SU(2)\times U(1))$ Lagrangian no extra sources of custodial breaking (besides the SM ones) are introduced, these operators appear at tree-level in the low-energy effective Lagrangian.  

\boldmath
\subsection{Spontaneous $SU(3)/(SU(2)\times U(1))$ symmetry breaking setup}
\unboldmath

An appropriate choice for the vacuum that breaks $SU(3)\to SU(2) \times U(1)$ is given by the following hermitian and orthogonal matrix:
\beq
\Delta_0 = 
\begin{pmatrix}
 \unity_2 & 0 \\
 0 & -1 \\
\end{pmatrix}\,,
\eeq
that satisfies the relations in Eq.~(\ref{Delta0Prop}). As in the previous cases, it is then possible to describe the massless excitations around the vacuum with a unitary field $\Delta(x)$ obtained ``rotating'' the vacuum with the GB non-linear field $\Omega(x)$:
\beq
\Delta(x)= \Omega(x)\, \Delta_0\, \Omega(x)^\dagger\,,\qquad\qquad
\Delta(x)\rightarrow \gtt\, \Delta(x)\, \gtt^\dagger\,. 
\eeq
As the vacuum is invariant under $SU(2)\times U(1)$ transformations, $\Delta(x)$ belongs to the adjoint of $SU(3)$.
Being $dim(SU(3)/(SU(2)\times U(1))) =4$, the field $\Delta(x)$ describes the dynamics of only four GBs, which will be then identified with the longitudinal components of the SM gauge bosons and the physical Higgs particle. Using the following relations between the vacuum $\Delta_0$ and the broken and unbroken generators,
\bea
\Delta_0 \, T_a \, \Delta_0 = T_a \,,\qquad \qquad \Delta_0 \, X_{\hat{a}}\,\Delta_0 = -X_{\hat{a}}\,,
\eea
and because of the relations in Eq.~(\ref{Delta0Prop}), the excitations around the vacuum can be arranged in the $\Sigma$-representation as in Eq.~(\ref{DeltaAndTheta}) with $\Omega$ and $\SH$ given as in Eq.~(\ref{OmegaThetaSO5}).
Choosing the following direction of EW symmetry breaking,
\beq
X= \frac{1}{\sqrt{2}} 
\begin{pmatrix}
0     &  e_2 \\  
e_2^T & 0  \\
\end{pmatrix}\,,
\eeq
it is possible to write the $SU(3)$ embedding of the SM GB fields as
\beq
\cX(x) = 
\sqrt{2} \left(
\begin{array}{cc}
\UH(x)  &   \\
  &  1 \\
\end{array}
\right) X \left(
\begin{array}{cc}
\UH(x)^\dag  &  \\
  &  1 \\
\end{array}\right) =  
\begin{pmatrix}
 0 & \UH(x) e_2  \\
 (\UH(x) e_2)^\dagger & 0  \\
\end{pmatrix}\,, 
\eeq
reducing to $\cX = \sqrt{2} X$ in the unitary gauge. As for the two  models previously analysed, the GB field matrix $\SH$ can be expressed in terms of $\cX$ as in Eq.~(\ref{ThetaSU5bis}). Finally the $SU(3)$-embedding of the $SU(2)_L\times U(1)_Y$ generators are given by
\beq
Q^a_L=\dfrac{1}{2}\left(
\begin{array}{cc}
 \sigma_a &    \\
  &  0 \\
\end{array}
\right),
\qquad\qquad
Q_Y=\dfrac{1}{6}\left(
\begin{array}{cc}
 \unity_2 &    \\
  &  -2 \\
\end{array}
\right)\,,
\eeq
with $\tr(Q^a_L Q^a_L)= 1$ and $\tr(Q_Y Q_Y)= 1/6$.

\boldmath
\subsection{The low-energy effective EW chiral Lagrangian}
\unboldmath 

By substituting the explicit expressions for $\SH$, $\VLt_\mu$, $\WLt_\mu$ and $\BLt_\mu$ into the operators of the high-energy basis in Eq.~(\ref{AppelquistBasisG}), $\LL_{\text{low}}$  is obtained for the $SU(3)/(SU(2)\times U(1))$ model as a function of the SM would-be GBs and the light physical Higgs $\varphi$.

\boldmath
\subsubsection{The two-derivative low-energy projection}
\unboldmath

The  low-energy projection of this CH model, where the custodial symmetry is not approximately conserved, underlines some peculiarities that can be already  seen in the resulting expression for the dimension-two operator $\cAt_{C}$:
\beq
\cAt_{C}=-\dfrac{f^2}{4}\Tr(\VLt_\mu\VLt^\mu) = \cP_H \,+\,\frac{4}{\xi}\sin^2\alfm \cP_C \,+\,\frac{2}{\xi}\sin^4\alfm \cP_T\,.
\label{ACsu3}
\eeq
It projects at low-energy not only into the $h$ and GBs kinetic terms as expected, but also into the two-derivative custodial violating operator $\cP_T$ in Eq.~(\ref{ExtALFOp2}). 

Alike to the situation for the models previously studied, $\cAt_{C}$ contains the term that describes the masses of the gauge bosons once the EW symmetry is broken. Requiring consistency with the definition of the $W$-mass, the link given in Eq.~(\ref{xidefSU5}) among the EW scale $v$, the Higgs VEV $\vh$ and the strong dynamic scale $f$ also  follows here.

\boldmath
\subsubsection{The four-derivative low-energy projection}
\unboldmath

The low-energy projection of the four-derivative operators  listed in Eq.~(\ref{AppelquistBasisG}) results in the following decomposition for the $SU(3)/(SU(2)\times U(1))$ model:
\begin{align}
\cAt_B =& \frac{2}{3}\cP_B\,, \nn \\
\cAt_W =& \cP_W\,, \nn \\
\cAt_{B\Sigma} =& -\frac{g^{\prime2}}{6} \left(1 + 3\cos \alfd \right) \cP_B\,, \nn \\
\cAt_{W\Sigma} =& -2\,g^2\,\cos \alf \cP_W + \sin^4 \alfm \cP_{12}\,, \nn \\
\cAt_1 =& \frac{1}{4}\sin^2\alf\cP_1 \,, \nn \\
\cAt_2 =& \frac{1}{4}\sin^2\alf\cP_2 +\dfrac{\sqrt{\xi}}{2}\sin\alfd\cP_4 \,, \nn \\
\cAt_3 =& \frac{1}{2}\sin^2\alf\cP_3 - 2\sqrt{\xi}\sin\alf\cP_5 + 2\sin^4\alfm \cP_{13} + 2\sqrt{\xi}\sin\alf\sin^2\alfm\cP_{17} \,, \nn \\
\cAt_4 =& 4\,\xi^2\,\cP_{DH} + 16\sin^4\alfm\cP_{6} - 16\,\xi\sin^2\alfm\cP_{20} + 8\,\xi\sin^4\alfm\cP_{21}+  \nn \\
& - 16\sin^6\alfm\cP_{23} + 4\sin^8\alfm \cP_{26} \,, \nn \\
\cAt_{5} =&4\,\xi^2\,\cP_{DH}  - 16\,\xi\sin^2\alfm\cP_{8} + 16\sin^4\alfm\cP_{11} + 8\,\xi\sin^4\alfm\cP_{22}  +\nn\\
&- 16\sin^6\alfm\cP_{24}+ 4\sin^8\alfm \cP_{26} \,,\label{ACsu3n4}\\
\cAt_{6} =& - 2\,\xi\,\cP_{\Box h} - \frac{1}{2}\sin^2\alf\cP_{6} - 2\sqrt{\xi}\sin\alf\left(\cP_{7}-2\cP_{10} \right) +
                4\xi \cos^2\alfm \cP_{8} + \nn \\
  &+ 4\sin^2\alfm \cP_{9} - 2\sin^4\alfm \left(\cP_{15} - 2\cP_{16}\right) - 2\xi \left(1+2\cos\alf\right) \sin^2\alfm \cP_{22}+  \nn\\
  &+ 2\sqrt{\xi}\sin\alf\sin^2\alfm \left(\cP_{18}-2\cP_{19}+\cP_{25}\right) + \sin^2\alf\sin^2\alfm\cP_{23} + \nn \\
  &- 4\sin^6\alfm \cP_{24}+ 2 \sin^8\alfm\cP_{26} \,, \nn \\
\cAt_{7} =& 2\,\xi^2\,\cP_{DH} + 8\sin^4\alfm \cP_6 - 4\xi\sin^2\alfm\cP_8 -  
      2\sqrt{\xi} \sin\alf \sin^2 \alfm \cP_{18} +      \nn \\
  &- 4\xi\sin^2\alfm \cP_{20} - 2 \xi\cos\alf\sin^2\alfm\cP_{21} + 2\xi\sin^2\alfm\cP_{22} - \nn \\
  &- 2\left(3-\cos\alf\right)\sin^4\alfm \cP_{23} + \sin^2\alf\sin^2\alfm\cP_{24} + 2\sin\alfm^8 \cP_{26} \nn 
\end{align}
The remaining operator in the list in Eq.~(\ref{AppelquistBasisG}) is not independent in this case, as it can be expressed as the combination
\beq
\cAt_{8} = \frac{1}{2}\cAt_{4} +\cAt_{5}-2\cAt_7\,,
\eeq
which in summary implies that the low-energy physical consequences of this model depend on nine arbitrary coefficients.

%
%
\section{Matching the high- and the low-energy Lagrangians}
\label{Sect:Matching}
 
The remnant of the GB nature of the Higgs field can be tracked down to the trigonometric functions that enter into the low-energy  EW chiral Lagrangian for the specific CH models: indeed, one given gauge vertex can involve an arbitrary number of $h$ legs, with a suppression in terms of powers of the GB scale $f$. The explicit dependence on the $h$ field is easily recovered using Eq.~(\ref{varphiExp}) in combination with trigonometric function properties.
In the general $\LL_{low}$ basis, the dependence on the $h$ field  is encoded into the generic functions $\cF_i(h)$  in Eq.~(\ref{LALF}) and into some operators which contain derivatives of $h$. The matching between the low-energy EW chiral Lagrangian of the specific CH models and the general $\LL_{low}$ basis  in Eq.~(\ref{LALFgen}) allows to identify the products $c_i\,\cF_i(h)$ in terms of the high-energy parameters. The existence of peculiar correlations between the low-energy chiral effective operators could indeed provide very valuable information when trying to unveil the nature of the EWSB mechanism~\cite{Brivio:2013pma,Brivio:2014pfa,Gavela:2014vra}.

\boldmath
\subsection{The $SU(5)/SO(5)$ and $SO(5)/SO(4)$ models}
\unboldmath

For the specific case of the $SU(5)/SO(5)$ model discussed in Sect.~\ref{Sect:GKmodel} and the terms in its two-derivative Lagrangian it results
\beq
\cF_C(h)=\frac{4}{\xi}\sin^2\left[\dfrac{\varphi}{2f}\right]\,,\qquad\qquad \cF_H(h)=1\,, 
\label{FCSU5SO5}
\eeq 
for the custodial preserving sector, while 
\beq
c_T\cF_T(h)=0\,
\eeq
for the custodial breaking term, as expected from a model which was formulated with an embedded custodial symmetry. 
 
A superficial look to the $\xi$ dependence of the right-hand side of Eq.~(\ref{LSU5n2}) (or equivalently of $\cF_C(h)$ in Eq.~(\ref{FCSU5SO5})) may raise questions about an apparent unphysical behaviour for $\xi \ll 1$. However, this is not the case as for $\xi \to 0$ Eq.~(\ref{LSU5n2}) reduces to
\beq
\cAt_{C}\approx \left[ \left(1+\frac{h}{v}\right)^2 - \frac{\xi}{12}\frac{h}{v}\left(1+\frac{h}{v}\right)\left(3+3\frac{h}{v}+\frac{h^2}{v^2}\right)+ \cO(\xi^2) \right]\cP_C \,+\, \cP_H \,,
\label{LSU5n2exp} 
\eeq
with the SM gauge boson-Higgs couplings exactly recovered\footnote{Equivalently,  rewriting $\cF_C(h)$ in Eq.~(\ref{FCSU5SO5}) as $\cF_C(h)= (\varphi^2/v^2) [\sin(x)/x]^2$ with $x \equiv \sqrt{\xi} \varphi/(2v)$ shows that its $\xi \rightarrow 0$ limit is safe, as $\sin(x)/x$ is an analytic function for any value of $x$ and in particular $x=0$.
} for $\xi=0$. This is consistent, as in this model the three would-be SM GBs and the Higgs field were introduced in a $SU(2)_L$ doublet structure embedded into the $SU(5)$ representation (see Eq.~(\ref{SU5higgs})). Any deviation from the SM (doublet) predictions should thus appear weighted by powers of $\xi$. For completeness, it may be useful to provide the expression for the $\cF_C(h)$ function in the notation usually adopted in the literature\footnote{In Ref.~\cite{Contino:2011np}, a slightly different result is reported: $a_C=1-\xi/2$ and $b_C=1-2\xi$. This is due to a different normalisation chosen for the operator $\cAt_C$ in Ref.~\cite{Contino:2011np} (see Eq.~(\ref{LinkAmongUsAndContinoFirst})). By a redefinition of the Higgs field, $\varphi\to\varphi/2$, the two expressions for $a_C$ and $b_C$ coincide.}:
\beq
\cF_C(h)=1+2a_C\frac{h}{v}+b_C\frac{h^2}{v^2}+\ldots\,,\qquad\text{with}\qquad
a_C=1-\frac{\xi}{8}\,,\quad
b_C=1-\frac{\xi}{2}\,.
\eeq
For the terms in the four-derivative Lagrangian, the expressions for the products $c_i\,\cF_i(h)$ are reported in Tab.~\ref{tableCustConserving} (second column).
\begin{table}[h!]
{\small
\renewcommand{\arraystretch}{2}
\hspace*{-1.4cm}
\begin{tabular}{|>{$}c<{$}|*3{>{$}c<{$}}|}
\hline
c_i\cF_i(h)& \parbox{2cm}{$SU(5)/SO(5)$\\$SO(5)/SO(4)$} & SU(3)/SU(2)\times U(1) &	\text{linear }d\leq6\\[1.7mm] 
\hline
\cF_C(h)
&\frac{4}{\xi}\sin^2\frac{\varphi}{2f}
&\frac{4}{\xi}\sin^2\frac{\varphi}{2f}
&1+\frac{(v+h)^2}{2\Lambda^2}c_{\Phi4}
\\
\cF_H(h)
&1
&1
&1+\frac{(v+h)^2}{2\Lambda^2}\left(c_{\Phi1}+2c_{\Phi2}+c_{\Phi4}\right)
\\
\cF_B(h)
&1-4g^{\prime 2}\ct_{B\Sigma}\cos^2\frac{\varphi}{2f}
&1- g'^2\frac{\ct_{B\Sigma}}{6}\left(1+3\cos\frac{2\varphi}{f}\right)
&1+\frac{(v+h)^2}{2\Lambda^2}g^{\prime 2}c_{BB}
\\
\cF_W(h)
&	1-4g^2\ct_{W\Sigma}\cos^2\frac{\varphi}{2f}
&1-2g^2\ct_{W\Sigma}\cos\frac{\varphi}{f}
&1+\frac{(v+h)^2}{2\Lambda^2}g^2c_{WW}
\\
c_{\square H}\cF_{\square H}(h)
&-2\ct_6\xi
&-2\ct_6\xi
&\frac{v^2}{2\Lambda^2}c_{\square\Phi}
\\
c_{\Delta H}\cF_{\Delta H}(h)
&-
&-
&-
\\
c_{D H}\cF_{D H}(h)
&4\left(\ct_4+\ct_5\right)\xi^2 
&2\left(2\ct_4+2\ct_5+\ct_7\right)\xi^2
&-
\\
c_1\cF_1(h)
&\ct_1\sin^2\frac{\varphi}{2f}
&\frac{\ct_1}{4}\sin^2\frac{\varphi}{f}
&\frac{(v+h)^2}{4\Lambda^2}c_{BW}
\\
c_2\cF_2(h)
&\ct_2\sin^2\frac{\varphi}{2f}
&\frac{\ct_2}{4}\sin^2\frac{\varphi}{f}
&\frac{(v+h)^2}{8\Lambda^2}c_B
\\
c_3\cF_3(h)
&2\ct_3\sin^2\frac{\varphi}{2f}
&\frac{\ct_3}{2} \sin^2\frac{\varphi}{f}
&\frac{(v+h)^2}{8\Lambda^2}c_W
\\
c_4\cF_4(h)
&\ct_2\sqrt\xi\sin\frac{\varphi}{f}
&\dfrac{\ct_2}{2}\sqrt\xi\sin\frac{2\varphi}{f}
&\frac{v(v+h)}{2\Lambda^2}c_B
\\
c_5\cF_5(h)
&-2\ct_3\sqrt\xi\sin\frac{\varphi}{f}
&-2\ct_3\sqrt\xi\sin\frac{\varphi}{f}
&-\frac{v(v+h)}{2\Lambda^2}c_{W}
\\
c_6\cF_6(h)
&16\ct_4\sin^4\frac{\varphi}{2f}-\dfrac{1}{2}\ct_6\sin^2\frac{\varphi}{f} 
&8(2\ct_4+\ct_7)\sin^4\frac{\varphi}{2f}-\dfrac{1}{2}\ct_6\sin^2\frac{\varphi}{f}
&\frac{(v+h)^2}{8\Lambda^2}c_{\square\Phi}
\\
c_7\cF_7(h)
&-2\ct_6\sqrt{\xi}\sin\frac{\varphi}{f} 
&-2\ct_6\sqrt{\xi}\sin\frac{\varphi}{f}
&\frac{v(v+h)}{2\Lambda^2}c_{\square\Phi}
\\
c_8\cF_8(h)
&-16\ct_5\xi\sin^2\frac{\varphi}{2f}+4\ct_6\xi\cos^2\frac{\varphi}{2f} 
&-4(4\ct_5+\ct_7)\xi\sin^2\frac{\varphi}{2f}+4\ct_6\xi\cos^2\frac{\varphi}{2f}
&-\frac{v^2}{\Lambda^2}c_{\square\Phi}
\\
c_9\cF_9(h)
&4\ct_6\sin^2\frac{\varphi}{2f} 
&4\ct_6\sin^2\frac{\varphi}{2f}
&-\frac{(v+h)^2}{4\Lambda^2}c_{\square\Phi}
\\
c_{10}\cF_{10}(h )
&4\ct_6\sqrt{\xi}\sin\frac{\varphi}{f} 
&4\ct_6\sqrt{\xi}\sin\frac{\varphi}{f}
&-\frac{v(v+h)}{\Lambda^2}c_{\square\Phi}
\\ 
c_{11}\cF_{11}(h)
&16\ct_5\sin^4\frac{\varphi}{2f} 
&16\ct_5\sin^4\frac{\varphi}{2f}
&-
\\
c_{20}\cF_{20}(h)
&-16\ct_4\xi\sin^2\frac{\varphi}{2f} 
&-4(4\ct_4+\ct_7)\xi\sin^2\frac{\varphi}{2f}
&-
\\\hline
\end{tabular}}
\caption{\em \small Expressions for the products $c_i\,\cF_i(h)$ of custodial preserving operators:   $SU(5)/SO(5)$ and $SO(5)/SO(4)$  in the second column,   $SU(3)/(SU(2)\times U(1))$  in the third column,  and the $d=6$ effective linear Lagrangian in the fourth column.  The ``-'' entries indicate no leading order contributions at low-energy to the corresponding operator. Notice that the kinetic terms are not canonically normalised at this stage.}
\label{tableCustConserving}
\end{table}
Some relevant conclusions can be inferred from these results:
\begin{itemize}
\item[i)] All custodial preserving operators entering the low-energy Lagrangian $\LL_\text{low}$, Eq.~(\ref{LALF}), are generated from the high-energy one $\LL_{\text{high}}$ for the $SU(5)/SO(5)$ CH model, with the exception of the operator $\cP_{\Delta H}$ in Eq.~(\ref{ExtALFOph2}) and $\cF_H(h)$, $\cF_{\square H}$ and $\cF_{D H}$. They cannot be originated due to the GB nature of the $\varphi$ field in that model, which forbids couplings with an odd number of Goldstone bosons, plus the fact that the departure from a pure Goldstone boson nature is through its vev $\langle \varphi\rangle \ne 0$, and not from any source containing derivatives~\footnote{Even when fermions will be considered explicitly, it is not expected to result in $\cP_{\Delta H}$  generated at low energies; but the additional fermionic operators expected could be rewritten in terms of $\cP_{\Delta H}$ (and other operators) via EOM. A similar reasoning applies to $\cF_H(h)$, $\cF_{\square H}$ and $\cF_{D H}$. We thus keep them here for 
generality.}.
\item[ii)] All other operators present in $\LL^{p^4}_\text{low}$ in Eq.~(\ref{LALF}) and not appearing in Tab.~\ref{tableCustConserving} describe effects of tree-level custodial breaking beyond the SM ones, and are thus absent in the low-energy $SU(5)/SO(5)$ effective chiral Lagrangian discussed.
\item[iii)] The arbitrary functions $\cF_i(h)$ of the generic  low-energy effective chiral Lagrangian $\LL_\text{low}$ in Eq.~(\ref{LALF}) become now a constrained set. 
Having chosen a specific CH model  reduces the number of free parameters in $\LL_\text{low}$: sixteen low-energy generic parameters contained in $c_i\cF_i(h)$ are now described in terms of the eight high-energy parameters $\ct_i$. 
\end{itemize}

As the EW chiral Lagrangian for the minimal $SO(5)/SO(4)$ model is the same of the one for the $SU(5)/SO(5)$ model, the results presented here also apply to the minimal $SO(5)/SO(4)$ model.

\boldmath
\subsection{The $SU(3)/(SU(2)\times U(1))$ model}
\unboldmath

The $\cF_{C}(h)$ and $\cF_H(h)$ functions  of the two-derivative low-energy chiral Lagrangian Eq.~(\ref{LALF}) stemming  from the high-energy $SU(3)/(SU(2)\times U(1))$ model turn out to be
\beq
\cF_C(h)=\frac{4}{\xi}\sin^2\left[\dfrac{\varphi}{2f}\right]\,,\qquad\qquad \cF_H(h)=1\,, 
\label{FCSU3}
\eeq
for the custodial preserving sector, and thus equal to that for $SU(5)/SO(5)$ and $SO(5)/SO(4)$ in Eq.~(\ref{FCSU5SO5}). This suggests that they are universal for composite models in which the Higgs is embedded as a $SU(2)_L$ doublet. 
For the custodial breaking sector, instead, it results 
\beq
c_T \cF_T(h)=\frac{2}{\xi}\sin^4\left[\dfrac{\varphi}{2f}\right]\,,
\eeq
and in this case the coefficient $c_T$ is not a free parameter, but is fixed by the high-energy operator $\cAt_C$. In consequence,  the experimental bounds on the $T$ parameter~\cite{Beringer:1900zz} translate into strong constraints on the parameter $\xi$ and on the strong dynamics scale $f$:
\beq
\alpha_\text{em}\Delta T=\dfrac{\xi}{4}  
\qquad\Longrightarrow\qquad
\xi\lesssim 0.014\,,\qquad f\gtrsim 2\TeV\,.
\eeq

\begin{table}[t!]
\hspace*{-1.5cm}
\renewcommand{\arraystretch}{2}
 \begin{tabular}{
 |*2{>{$}c<{$}|}|
 *2{>{$}c<{$}|}
 }
 \hline
c_i \cF_i(h)& SU(3)/(SU(2)\times U(1)) & 
c_i \cF_i(h)& SU(3)/(SU(2)\times U(1)) \\
\hline
c_T\cF_T(h)
&\frac{2}{\xi}\sin^4\frac{\varphi}{2f}
&c_{21}\cF_{21}(h)
&8\ct_4\xi\sin^4\frac{\varphi}{2f}-2\ct_7\xi\cos\frac{\varphi}{f}\sin^2\frac{\varphi}{2f}
\\
c_{12}\cF_{12}(h)
& \ct_{W\Sigma}\sin^4\frac{\varphi}{2f}
&c_{22}\cF_{22}(h)
&8\ct_5\xi\sin^4\frac{\varphi}{2f}+2\xi\ct_7\sin^2\frac{\varphi}{2f}-2\ct_6\xi\sin^2\frac{\varphi}{2f}\left(1+2\cos\frac{\varphi}{f}\right)
\\
c_{13}\cF_{13}(h)
&2\ct_3\sin^4\frac{\varphi}{2f}
&c_{23}\cF_{23}(h)
&\multirow{2}{8cm}{\centering $-16\ct_4\sin^6\frac{\varphi}{2f}+\ct_6\sin^2\frac{\varphi}{2f}\sin^2\frac{\varphi}{f}
+2\ct_7 \sin ^4\frac{\varphi}{2f} \left(\cos\frac{\varphi}{f}-3\right)$}
\\
c_{15}\cF_{15}(h)
&-2\ct_6\sin^4\frac{\varphi}{2f}
&&
\\
c_{16}\cF_{16}(h)
&4\ct_6\sin^4\frac{\varphi}{2f}
&c_{24}\cF_{24}(h)
&-4(4\ct_5+\ct_6)\sin^6\frac{\varphi}{2f}+\ct_7\sin^2\frac{\varphi}{2f}\sin^2\frac{\varphi}{f} 
\\
c_{17}\cF_{17}(h)
&2\ct_3\sqrt\xi\sin^2\frac{\varphi}{2f}\sin\frac{\varphi}{f}
&c_{25}\cF_{25}(h)
&2\ct_6\sqrt\xi\sin^2\frac{\varphi}{2f}\sin\frac{\varphi}{f}
\\
c_{18}\cF_{18}(h)
&2(\ct_6-\ct_7)\sqrt\xi\sin^2\frac{\varphi}{2f}\sin\frac{\varphi}{f}
&c_{26}\cF_{26}(h)
&2(2(\ct_4+\ct_5)+\ct_6+\ct_7)\sin^8\frac{\varphi}{2f}
\\
c_{19}\cF_{19}(h)
&-4\ct_6\sqrt\xi\sin^2\frac{\varphi}{2f}\sin\frac{\varphi}{f}
&
&
\\
\hline
\end{tabular}
\caption{\em \small Expressions for the products $c_i\,\cF_i(h)$ for the custodial symmetry breaking operators of $SU(3)/(SU(2)\times U(1))$ CH model. No analogous contributions are present neither for the $SU(5)/SO(5)$ and  $SO(5)/SO(4)$ model, nor for the linear $d=6$ effective Lagrangian, but for the combination $c_T\cF_T(h)$ that receives contributions from $\cO_{\Phi1}$. }
\label{tableCustBreaking}
\end{table}

For the terms in the four-derivative Lagrangian, the expressions for the products $c_i\,\cF_i(h)$ corresponding to custodial invariant operators are reported in Tab.~\ref{tableCustConserving} (third column), while those corresponding to custodial-breaking ones are  collected in Tab.~\ref{tableCustBreaking}.

Contrary to the case of the two models previously analysed, all custodial preserving and all custodial breaking operators  entering the low-energy Lagrangian $\LL_\text{low}$ in Eq.~(\ref{LALF}) are generated from the high-energy one for the $SU(3)/(SU(2)\times U(1))$ CH model, with the exception of the operator $\cP_{\Delta H}$ in Eq.~(\ref{ExtALFOph2}) and $\cF_H(h)$, $\cF_{\square H}$ and $\cF_{D H}$. On the other side, also in this case the  a priori many arbitrary combinations $c_i\cF_i(h)$ can be written in terms of the small set of nine high-energy parameters $\ct_i$.\\

In summary, a quite universal pattern is suggested by our results as to the form of the $c_i\cF_i(h)$ functions, at least for the custodial preserving sector. Tab.~\ref{tableCustConserving} encompasses the main results and allows a direct comparison of the low-energy impact of the models considered (as well as of the BSM physics expected from linear realisations of EWSB). Not only $\cF_C(h)$  coincides exactly for all three chiral models considered, see Eqs.~(\ref{FCSU5SO5}) and (\ref{FCSU3}), but the $c_i\cF_i(h)$  functions for {\it all} four-derivative chiral operators do as well, except for the couplings which involve gauge field-strengths for which the intrinsically custodial-invariant groups and $SU(3)/(SU(2)\times U(1))$  differ simply by a rescaling of the scale $f$ and multiplicative factors, see Tab.~\ref{tableCustConserving}. 

\boldmath
\subsection{The $\xi \ll 1$ limit and the linear effective Lagrangian}
\unboldmath

As anticipated in Sect.~\ref{Sect:ExtendedALF}, the low-energy effective Lagrangian $\LL_{\text{low}}$  is suitable to describe a large class of Higgs models, including the case of a linearly realised EWSB. In the limit of small $\xi$, the trigonometric functions containing the Higgs field $\varphi$ can be expanded in Taylor series. If only the first terms in this expansion are retained, the resulting effective chiral Lagrangian describes similar interactions as the effective $d=6$ linear Lagrangian~\cite{Buchmuller:1985jz,Grzadkowski:2010es} -- and with similar features. For definiteness, let us refer to a specific basis for the bosonic sector of the effective $d=6$ linear Lagrangian  -- the so-called Hagiwara-Ishihara-Szalapski-Zeppenfeld (HISZ) basis~\cite{Hagiwara:1993ck,Hagiwara:1996kf}. The effective linear Lagrangian including the leading corrections can be decomposed as the SM part  plus a piece containing operators with canonical dimension $d=6$,  weighted down by suitable powers of the ultraviolet 
cut-off scale $\Lambda$:
\beq
\LL_\text{linear} = \LL_{SM}+\Delta\LL_\text{linear}\,,
\label{Llinear}
\eeq
where 
\beq
\Delta \LL_\text{linear} = \sum_i \frac{c_i}{\Lambda^2} \cO_i\,, 
\label{DeltaLlinear}
\eeq
with $c_i$ being order one parameters and $\cO_i$ denoting operators defined as follows~\cite{Hagiwara:1993ck,Hagiwara:1996kf,Grzadkowski:2010es}:
\beq
\begin{aligned}
&\cO_{BB} = \Phi^{\dagger} \hat{B}_{\mu \nu} \hat{B}^{\mu \nu} \Phi\,,
&&\cO_{WW} = \Phi^{\dagger} \hat{W}_{\mu \nu} \hat{W}^{\mu \nu} \Phi\,, \\
&\cO_W  = (\DL_{\mu} \Phi)^{\dagger} \hat{W}^{\mu \nu}  (\DL_{\nu} \Phi)\,, 
&&\cO_{BW} =  \Phi^{\dagger} \hat{B}_{\mu \nu} \hat{W}^{\mu \nu} \Phi\,, \\
&\cO_B  =  (\DL_{\mu} \Phi)^{\dagger} \hat{B}^{\mu \nu}  (\DL_{\nu} \Phi)\,,
&&\cO_{\Phi,1} =  \left (\DL_\mu \Phi \right)^\dagger \Phi\  \Phi^\dagger \left (\DL^\mu \Phi \right )\,,\\
&\cO_{\Phi,2} = \frac{1}{2} \partial^\mu\left ( \Phi^\dagger \Phi \right)
\partial_\mu\left ( \Phi^\dagger \Phi \right)\,,
&&\cO_{\Phi,3}=\dfrac{1}{3}\left(\Phi^\dag\Phi\right)^3\,,\\
&\cO_{\Phi,4} = \left (\DL_\mu \Phi \right)^\dagger \left(\DL^\mu\Phi\right)\left(\Phi^\dagger\Phi \right )\,,
&&\cO_{\square \Phi}=\left(\DL_\mu \DL^\mu\Phi\right)^\dag\left(\DL_\nu \DL^\nu\Phi\right)\,,
\end{aligned}
\label{LinearOpsGauge} 
\eeq
with $\DL_\mu\Phi\equiv \left(\partial_\mu+ \frac{i}{2} g' B_\mu + \frac{i}{2}g\sigma_i W^i_\mu \right)\Phi $.  Among these, $\cO_{\Phi,1}$ is custodial breaking and $\cO_{\Phi,3}$ is a pure potential-like Higgs term; assuming custodial symmetry it remains a total of eight independent operators\footnote{The original HISZ basis includes in addition the gluonic operator $\cO_{GG} = \Phi^\dagger \Phi \,G^a_{\mu\nu} G^{a\mu\nu}$, which is not considered here as only the EW sector is analysed in this paper.}.

After  the $SU(2)_L$ Higgs doublet $\Phi$ acquires a vev, $\mean{\Phi}= (v+h)/\sqrt{2}$, the interactions resulting from this set of linear operators can be also described by $\LL_{\text{low}}$ in Eq.~(\ref{LALF}), with the products $c_i\,\cF_i(h)$ taking the values shown in the last column of Tab.~\ref{tableCustConserving}. 
In the small $\xi$ limit, the low-energy effective chiral Lagrangian associated to the considered CH models converges to the linear one, with the correspondence
\begin{equation}
\ct_{B\Sigma}\to c_{BB}\,,\quad
\ct_{W\Sigma}\to c_{WW}\,,\quad
\ct_1\to c_{BW}\,,\quad
\ct_2\to c_{B}\,,\quad
\ct_3\to c_{W}\,,\quad
\ct_6\to c_{\square\Phi}\,.
\end{equation}
The parameters $\ct_4$ and $\ct_5$ are not relevant, because they appear in contributions of order $\xi^{\geq2}$, that correspond to linear operators of $d\geq8$. 
Notice in addition that the products $c_i\cF_i(h)$ corresponding to custodial-breaking operators and appearing in Tab.~\ref{tableCustBreaking} are suppressed by $\xi^{\ge2}$ and are therefore negligible in the small $\xi$ limit. Consistently, the corresponding contributions from the effective linear Lagrangian come from operators with dimensions $d\ge8$. Notice that a complete comparison is only possible in the basis where the kinetic terms are canonical: in Tab.~\ref{tableCustConserving}, $\cF_H(h)$ is 1 for the CH models, but not for the linear Lagrangian.

Eq.~(\ref{ACsu3n4}) together with the  decomposition in  Eq.~(\ref{LEwithG}) allow to appreciate the coincidences and the differences between the low-energy effective chiral Lagrangian and the effective linear one: i)  the gauge interactions stemming from some chiral operators combine with fixed weights, even for large $\xi$ values, which are precisely those predicted by the linear Lagrangian (see Tab.~\ref{tableCustConserving}); this is because the $\varphi$ field was embedded  in the high energy theory as a $SU(2)_L$ doublet; ii) the low-energy differences stem from the $h$ dependence, given via functions of $\sin\left[(\vh+h)/2f\right]$  for the low-energy chiral Lagrangian versus  powers of $\left(v+h\right)/2$ for linear realisations of BSM theories. Therefore, although the number of free parameters is the same in the two Lagrangians, the $h$-couplings have different dependencies~\cite{Azatov:2012bz,Isidori:2013cga}.  

To illustrate it, consider the $\cO_B$ operator of the linear realisation. Expressing $\Phi$  in terms of the GB matrix $\UH$ and the physical scalar $h$,
\beq
\Phi=\dfrac{(v+h)}{\sqrt2}\UH
\left(
\begin{array}{c}
 0 \\
 1 \\
\end{array}
\right)\,,
\eeq
 $\cO_B$ can be rewritten in the chiral notation as
\beq
\begin{split}
\cO_B=&B_{\mu\nu}\tr(\TL[\VL^\mu,\VL^\nu]) \dfrac{(v+h)^2}{4}+B_{\mu\nu}\tr(\TL\VL^\mu)\derp^\nu \dfrac{(v+h)^2}{4}\\
=&\left(\cP_2+2\,\hat\cP_{4\nu}\,\derp^\nu\right)\,\dfrac{(v+h)^2}{4}\,,
\end{split}
\eeq
to be compared with $\cAt_2$ in Eq.~(\ref{LEwithG}). 
This pattern is general for the complete set of operators: same gauge couplings as in $d=6$ linear basis for a fixed number of $h$ legs,  while the relative strength of couplings involving different number of $h$ external legs differs from that in linear expansions.  The results support the approach to the effective Lagrangian for composite Higgs models based in the linear expansion in Ref.~\cite{Hagiwara:1993ck,Hagiwara:1996kf,Grzadkowski:2010es,Giudice:2007fh} only if the Higgs is assumed to be a pure $SU(2)_L$ doublet. Indeed in this case $\xi\ll1$ and the trigonometric dependence on $h$ reduces exactly to the linear one, as $\sin^2(\varphi/f)= \xi(1+h/v)^2+\cO(\xi^2)$ and the higher order terms in $\xi$ can be safely neglected. 

Promising discriminating signals include then some pure-gauge versus gauge-Higgs couplings~\cite{Brivio:2013pma,Brivio:2014pfa,Gavela:2014vra}, whose precise form we have determined here for the specific CH models considered. The strength of this type of departures from the SM expectations depends on $\xi$ and therefore the larger $\xi$ the sooner it will be possible to disentangle at colliders a composite from an elementary nature of the Higgs particle.
 
 For the more general case in which the observed light Higgs particle is not an exact $SU(2)_L$ doublet, linear $d=6$ expansions will be insufficient to describe the leading corrections. There are then more independent parameters, as given by the general low-energy non-linear Lagrangian~\cite{Alonso:2012px,Brivio:2013pma,Gavela:2014vra}, and further decorrelations are expected, including among vertices with the same number of Higgs legs.

%
%
\section{Conclusions}
\label{Sect:Conc}

For a simple group $\cG$ broken to a subgroup $\cH$, we have constructed the effective chiral Lagrangian for a generic symmetric coset $\cG/\cH$, restricting to CP-even bosonic operators with at most four derivatives: at most seven independent operators result, aside from the kinetic terms. After gauging the $SU(2)_L \times U(1)_Y$ symmetry and considering the induced custodial symmetry breaking terms, the total number of operators increases up to ten, plus three kinetic terms. This finding  is independent of the specific choice of $\cG$. It applies to composite Higgs scenarios in which the Higgs particle is a pseudo-Goldstone boson of the spontaneous breaking of $\cG$,  irrespective of the $SU(2)_L$ representation to which it may belong.

One consequence is that  for any  composite model in which the Higgs is embedded as a Goldstone boson of the high-energy theory, we predict strong relations among the dozens of low-energy parameters of the general low-energy effective chiral Lagrangian with a light Higgs particle.

Under the assumptions of no new sources of custodial non-invariance other than the SM gauge ones, we then particularised to the case of three specific composite Higgs models: two intrinsically custodial-preserving ones, $SU(5)/SO(5)$ and $SO(5)/SO(4)$,  and another which by construction breaks custodial symmetry, 
$SU(3)/(SU(2)\times U(1))$.  For the latter group the number of independent operators is nine (aside from possible sources of explicit subsequent breaking and from kinetic terms), while for the former two groups it is eight. 

 This analysis has allowed to confirm that the general low-energy Lagrangian for a dynamical Higgs particle developed in Refs~\cite{Alonso:2012px,Brivio:2013pma} is complete:  all operators  of that basis  and nothing else result at low-energies. The exceptions are $\cP_{\Delta H}$ in Eq.~(\ref{WBSU(5)}) and $\cF_H(h)$, $\cF_{\square H}$ and $\cF_{D H}$, which are not generated: these couplings are forbidden by the original Goldstone boson nature of the Higgs particle, and also by the particular way in which the global symmetry is subsequently explicitly broken in the models considered (as a vev for the Higgs particle). The results of the sigma decomposition confirm as well the powers of $\xi$ predicted in Ref.~\cite{Alonso:2012px,Brivio:2013pma} as  weights for each operator of the low-energy effective chiral Lagrangian, allowing an immediate comparison with linear expansions. Note that, for the scalar sector, a different and fully model-independent proof of the completeness of that effective 
Lagrangian  is provided by the recent analysis in Ref~\cite{GKMS} of one-loop induced four-derivative counterterms.

The present work also sheds light on the relevance of the Higgs particle being a Goldstone boson embedded as a part of an $SU(2)_L$ doublet in a representation of the high-energy group, versus scenarios in which it is also a Goldstone boson albeit a $SU(2)$ singlet, or the most general case in which  $h$ is a generic singlet scalar,  such as a Higgs ``impostor", or a dilaton or a dark sector scalar.  
Data strongly suggest that $h$ belongs to an electroweak doublet and it is thus especially interesting to further explore the consequences of this restriction for BSM physics. Our results show that:
\begin{itemize}
\item[i)] 
For vertices with a fixed number of external Higgs legs, the gauge couplings combine with the same relative weights as in the case of the $d=6$ linear  effective Lagrangian for BSM physics. This is so irrespective of the size of $\xi$ for the intrinsically custodial preserving groups considered, while for the $SU(3)/(SU(2)\times U(1))$ model it only holds at leading order in $\xi$.
\item[ii)] Conversely, vertices with different number of $h$ external legs get different relative weights than in linear realisations of BSM physics.  While the latter show a generic polynomial Higgs dependence on $(v+h)$ and its derivatives, composite Higgs models induce a functional $c_i\cF_i(h)$ dependence in the effective Lagrangian. We have explicitly determined all  $c_i\cF_i(h)$ functions of the low-energy effective Lagrangian up to four-derivative couplings, for the three composite models considered.  
\item[iii)]  The determined $\cF_i(h)$ are trigonometric functions, as befits a Goldstone boson origin of the Higgs field, and it is tantalising that they turn out to be basically exactly equal for the three models considered, except in the set of operators which include gauge field strengths; even the latter differ at most by a rescaling of $f$ (aside from custodial breaking ingredients). 
\end{itemize}
The latter point suggests that the  $\cF_i(h)$ determined here may be universal to any composite Higgs model. Tab.~\ref{tableCustConserving} encompasses the main results and allows a direct comparison of the low-energy impact of the composite Higgs models considered (as well as of BSM linear realisations of EWSB). This universality may be very relevant for the analysis of  experimental data,  as it predicts the precise form in which anomalies in Higgs-gauge couplings and self-couplings would point to composite Higgs models, and in an almost model-independent way.  

The present work  illuminates as well the relation between non-linear realisations of electroweak symmetry breaking with a light Higgs embedded as an electroweak doublet of the high-energy strong dynamics, and linear ones. The former approximated the latter when the strong dynamics scale grows, that is for $\xi\rightarrow0$. We have shown here that the precise -and almost universal-  $\cF_i(h)$  functions determined for three composite Higgs models shows the specific form of the convergence towards the Higgs dependence of linear realisations, in the limit $\xi\ll 1$.

If the Higgs particle is a Goldstone boson of the high-energy group, although not an electroweak doublet -for instance a singlet- then point i) above would not hold: while the number of arbitrary operator coefficients would still be restricted to the small number predicted for a generic symmetric coset,  the relative weight of gauge couplings for a fixed number of external $h$ legs would be different with respect to that in  linear analysis~\cite{Hagiwara:1993ck,Hagiwara:1996kf,Grzadkowski:2010es,Giudice:2007fh}, with gauge decorrelations predicted alike to those in Refs.~\cite{Alonso:2012px,Brivio:2013pma,Brivio:2014pfa}. 
Finally, for the completely general case in which the Higgs field is a generic SM scalar singlet at low energies, again the linear-based analysis is not an appropriate tool as both the relative weights of gauge couplings with and without the same number of $h$ legs are completely free parameters,  described (in the absence of a concrete model) by the most general low-energy bosonic effective Lagrangian for a dynamical Higgs~\cite{Alonso:2012px,Brivio:2013pma}.
Further experimental decorrelations and signals follow in these last two cases~\cite{Brivio:2013pma,Brivio:2014pfa}. It is particularly relevant to keep tracking the possible non-doublet components of the Higgs particle, in view of the present large error bars and the theoretical challenge set by the electroweak hierarchy problem.

\section*{Acknowledgements}
We acknowledge illuminating conversations with Roberto Contino, Ferruccio Feruglio, Howard Georgi, Concha Gonzalez-Garc\'ia, Christophe Grojean, Gino Isidori, Kirill Kanshin, Pedro Machado, Luciano Maiani, Aneesh Manohar, Sara Saa and Michael Trott. We also thank Oscar Cat\`a for a clarifying discussion on the issues risen in Ref.~\cite{Buchalla:2013rka}. We also acknowledge partial support of the European Union network FP7 ITN INVISIBLES (Marie Curie Actions, PITN-GA-2011-289442), of CiCYT through the project  FPA2012-31880, of the European Union FP7 ITN UNILHC (Marie Curie Actions, PITN-GA-2009-237920), of MICINN through the grant BES-2010-037869, of the Spanish MINECO's ``Centro de Excelencia Severo Ochoa'' Programme under grant SEV-2012-0249, and of the Italian Ministero dell'Uni\-ver\-si\-t\`a e della Ricerca Scientifica through the COFIN program (PRIN 2008) and the contract MRTN-CT-2006-035505. RA acknowledges support by the DOE grant DE-SC0009919. The work of I.B. is supported by an ESR contract of the European Union network FP7 ITN INVISIBLES mentioned above. The work of L.M. is supported by the Juan de 
la Cierva programme (JCI-2011-09244). I.B., B.G., L.M, and S.R. acknowledge the Galileo Galilei Institute (Florence) and CERN TH department for hospitality during the initial stages of this work.

%
%
\appendix
\section{The $\Omega$-representation}

The CCWZ construction allows to identify a non-redundant parametrisation of the GB fields arising from the breaking $\cG\to\cH$ in terms of the GB matrix $\Omega$, or in terms of $\SH$. Although the choice between the $\Omega$-representation or the $\Sigma$-representation is not discriminant to construct the most general effective Lagrangian for the $\cG/\cH$ coset, much of the CH models in the literature have been presented in the $\Omega$-representation. In the following, we will rewrite the effective Lagrangians in Eqs.~(\ref{GbasisV})-(\ref{GbasisVV}) and in Eq.~(\ref{LLG}) in the $\Omega$-representation and compare them with other effective Lagrangians presented in the literature for the case of $SO(5)/SO(4)$ model.

\subsection{The high-energy effective chiral Lagrangian}
\label{OurBasesOmegaRep}

The building blocks used to construct the effective Lagrangian in Eqs.~(\ref{GbasisV})-(\ref{GbasisVV}) are the gauge field strength $\SLt_{\mu\nu}$ and the vector chiral field $\VLt_\mu$, which in the $\Sigma$-representation transform in the adjoint of the group $\cG$ (see Eqs.~(\ref{TransS2}) and (\ref{newchiral}), respectively). To move to the $\Omega$-representation, it is then necessary to translate $\SLt_{\mu\nu}$ and $\VLt_\mu$, and their graded versions, into building blocks that transform in the adjoint of the preserved subgroup $\cH$. The GB matrix $\Omega$ can be exploited to this end:
\beq
\begin{aligned}
s_{\mu\nu}&\equiv \Omega^{-1}\, \SLt_{\mu\nu}\,\Omega\,,\qquad\qquad
&v_{\mu}&\equiv \Omega^{-1}\, \VLt_{\mu}\,\Omega=\Omega^{-1}\DL_\mu\Omega-\Omega\DL_\mu\Omega^{-1}\,,\\
s^\cR_{\mu\nu}&\equiv \Omega\, \SLt^\cR_{\mu\nu}\,\Omega^{-1}\,,\qquad\qquad
&v^\cR_{\mu}&\equiv \Omega\, \VLt^\cR_{\mu}\,\Omega^{-1}=\Omega\DL_\mu\Omega^{-1}-\Omega^{-1}\DL_\mu\Omega\,.
\end{aligned}
\label{OmegaBuildingBlocks}
\eeq
From the relation $v_{\mu}+v^\cR_{\mu}=0$, one can deduce that $v_\mu$ runs only over the broken generators and not over the preserved ones. It is then useful to introduce the following notation: 
\beq
\Omega^{-1} \DL_\mu \Omega \equiv \frac{v_\mu}{2} + i\, p_\mu = \frac{v^{\hat{a}}_\mu}{2} \,X_{\hat{a}} + i\, p^a_\mu \,T_a \,,
\label{masterCCWZus}
\eeq
with $v_\mu$ and $p_\mu$ transforming under $\cH$ as:
\beq
v_\mu\to \htt\,v_\mu\,\htt^{-1}\,,\qquad\qquad
p_\mu\to \htt\left(p_\mu-i \derp_\mu\right)\htt^{-1}\,.
\eeq
The field $p_\mu$ transforms as a connection and therefore it is possible to define the extended covariant derivative of $v_\mu$ as
\beq
\nabla_\mu v_\nu=\DL_\mu v_\nu+i[e_\mu,v_\nu]\,.
\eeq
From Eq.~(\ref{OmegaBuildingBlocks}), if follows that 
\beq
v^\cR_{\mu}=-v_{\mu}\,,
\eeq
and therefore the list of building blocks necessary to construct the effective Lagrangian in the $\Omega$-representation reduces to only three elements: $\{v_\mu,\,s_{\mu\nu},\, s^\cR_{\mu\nu}\}$.

The high-energy basis for a generic symmetric coset $\cG/\cH$ in the custodial preserving framework presented in Eqs.~(\ref{GbasisV})-(\ref{GbasisVV}) reads in the $\Omega$-representation: 
\beq
\begin{aligned}
\Tr\left(\VLt_\mu\VLt^\mu\right)&\rightarrow \Tr\left(v_\mu v^\mu\right) \,,\\
\Tr\left(\SLt_{\mu \nu}\SLt^{\mu \nu}\right)&\rightarrow \Tr\left(s_{\mu \nu}s^{\mu \nu}\right)\,,\\
\Tr\left(\SH\,\SLt^R_{\mu\nu}\,\SH^{-1}\,\SLt^{\mu\nu}\right)&\rightarrow \Tr\left(s^\cR_{\mu\nu}s^{\mu\nu}\right)\,,\\
\Tr\left(\SLt_{\mu\nu}\left[\VLt^\mu,\VLt^\nu\right]\right)&\rightarrow\Tr\left(s_{\mu\nu}\left[v^\mu,v^\nu\right]\right)\,,\\
\Tr\left(\VLt_\mu\,\VLt^\mu\right)\Tr\left(\VLt_\nu\,\VLt^\nu\right)&\rightarrow\Tr\left(v_\mu\,v^\mu\right)\Tr\left(v_\nu\,v^\nu\right)\,,\\
\Tr\left(\VLt_\mu\,\VLt_\nu\right) \Tr\left(\VLt^\mu\,\VLt^\nu\right)&\rightarrow\Tr\left(v_\mu\,v_\nu\right) \Tr\left(v^\mu\,v^\nu\right)\,, \\
\Tr\left((\cD_\mu\VLt^\mu)^2 \right)&\rightarrow\Tr\left((\nabla_\mu v^\mu)^2 \right)\,,\\
\Tr\left(\VLt_\mu\,\VLt^\mu\VLt_\nu\,\VLt^\nu\right)&\rightarrow\Tr\left(v_\mu\,v^\mu\,v_\nu\,v^\nu\right) \,,\\
\Tr\left(\VLt_\mu\,\VLt_\nu\VLt^\mu\,\VLt^\nu\right)&\rightarrow\Tr\left(v_\mu\,v_\nu\,v^\mu\,v^\nu\right)\,.
\end{aligned}
\label{BasisCPinOmega}
\eeq
\\

In realistic realisations of CH models only the SM gauge group is gauged, and in this case the previous basis is augmented by  operators constructed with 
\beq
b_{\mu\nu}\equiv \Omega^{-1}\, \BL_{\mu\nu}\,\Omega\,, 
\qquad \text{and} \qquad
w_{\mu\nu}\equiv \Omega^{-1}\, \WL_{\mu\nu}\,\Omega\,,
\eeq
in substitution of those containing explicit gauge field strength $s_{\mu\nu}$. As a result, the effective Lagrangian in Eq.~(\ref{LLG}) reads in the $\Omega$-representation:
\beq
\begin{aligned}
\cAt_C&\rightarrow-\frac{f^2}{4}\Tr\left(v_\mu v^\mu\right)\,,\qquad\qquad
&\cAt_3&\rightarrow g\Tr\left(w_{\mu\nu}\left[v^\mu,v^\nu\right]\right)\,,\\
\cAt_B&\rightarrow g^{\prime2}\Tr\left(b_{\mu \nu}b^{\mu \nu}\right)\,,\qquad\qquad
&\cAt_4&\rightarrow \Tr\left(v_\mu\,v^\mu\right)\Tr\left(v_\nu\,v^\nu\right)\,,\\
\cAt_W&\rightarrow g^2\Tr\left(w_{\mu \nu}w^{\mu \nu}\right)\,,\qquad\qquad
&\cAt_5&\rightarrow\Tr\left(v_\mu\,v_\nu\right) \Tr\left(v^\mu\,v^\nu\right)\,, \\
\cAt_{B\Sigma}&\rightarrow g^{\prime2}\Tr\left(b^\cR_{\mu \nu}b^{\mu \nu}\right)\,,\qquad\qquad
&\cAt_6&\rightarrow \Tr\left((\nabla_\mu v^\mu)^2 \right)\,,\\
\cAt_{W\Sigma}&\rightarrow g^2\Tr\left(w^\cR_{\mu \nu}w^{\mu \nu}\right)\,,\qquad\qquad
&\cAt_7&\rightarrow\Tr\left(v_\mu\,v^\mu\,v_\nu\,v^\nu\right) \,,\\
\cAt_1&\rightarrow g g'\Tr\left(b^\cR_{\mu \nu}w^{\mu \nu}\right)\,,\qquad\qquad
&\cAt_8&\rightarrow\Tr\left(v_\mu\,v_\nu\,v^\mu\,v^\nu\right)\,.\\
\cAt_2&\rightarrow g'\Tr\left(b_{\mu\nu}\left[v^\mu,v^\nu\right]\right)\,,
&
\end{aligned}
\label{BasisCBinOmega}
\eeq
where $b^\cR_{\mu\nu}\equiv \Omega\, \BL_{\mu\nu}\,\Omega^{-1} $  and $w^\cR_{\mu\nu}\equiv \Omega\, \WL_{\mu\nu}\,\Omega^{-1}$, and the gauge constants $g$ and $g'$ have been explicitly reported.

\boldmath
\subsection{$SO(5)/SO(4)$ model in the $\Omega$-representation by Refs.[49,57]}
\unboldmath
\label{App:Contino}

An effective Lagrangian for the $SO(5)/SO(4)$ CH model has been explicitly derived in the $\Omega$-representation in Refs.~\cite{Contino:2011np,Azatov:2013ura}, even if slightly different operator bases have been reported in the two articles. Furthermore, a different notation has been adopted in these references with respect to the notation used in this paper. In this section, we will comment on the differences among the two bases in Refs.~\cite{Contino:2011np,Azatov:2013ura}. In the next section, we will discuss the different notations used and compare between the operators basis in Ref.~\cite{Contino:2011np} and the one presented in appendix~\ref{OurBasesOmegaRep}.

Equivalently to the definition in Eq.~(\ref{masterCCWZus}), it is possible to introduce the following expression, according to Refs.~\cite{Contino:2011np,Azatov:2013ura},
\beq
-i\, U^{-1} \DL_\mu U \equiv d_\mu + e_\mu = d^{\hat{a}}_\mu \,X_{\hat{a}} + e^a_\mu \,T_a \,,
\label{ccwz2}
\eeq
where $U$ stands for the GB matrix of the $SO(5)/SO(4)$ coset, defined in Eq.~(11) of Ref.~\cite{Contino:2011np}, and $d_\mu$ and $e_\mu$ transform under a global transformation of $\cH=SO(4)$ as
\beq
d_\mu \raw \htt \,d^\mu \,\htt^{-1} \,, \qquad \qquad 
e_\mu \raw \htt \left(e_\mu -i\, \partial_\mu\right)\,\htt^{-1} \,.
\eeq
The field $e_\mu$ transforms as a connection, opening the possibility to define the field strength
\beq
e_{\mu\nu} \equiv \partial_\mu e_\nu - \partial_\nu e_\mu + i\left[e_\mu , e_\nu \right] \,,
\eeq
and  the extended covariant derivative of $d_\mu$ as
\beq
\nabla_\mu d_\nu = \DL_\mu d_\nu + i\left[e_\mu , d_\nu \right]\,,
\eeq
where the covariant derivative $\DL_\mu d_\nu$ is defined in terms of the gauge fields $F_\mu$ associated to the gauging of a subgroup $SO(4)'$ of $SO(5)$:
\beq
\DL_\mu d_\nu= \partial_\mu d_\nu + i\,g_S\,F_\mu\, d_\nu\,.
\eeq
It is then possible to introduce the $F_{\mu\nu}$ gauge field strength, albeit transforming in the adjoint of the group $SO(4)$:
\beq
\begin{aligned}
f_{\mu\nu} = \Omega^{-1} \, F_{\mu\nu} \Omega\,, \qquad\qquad 
f_{\mu\nu} \raw \htt\, f_{\mu\nu} \, \htt^{-1}\,,\\
f_{\mu\nu}^\cR = \Omega \, F_{\mu\nu}^\cR\Omega^{-1}\,, \qquad\qquad 
f^\cR_{\mu\nu} \raw \htt\, f^\cR_{\mu\nu} \, \htt^{-1} \,,
\end{aligned}
\eeq
where in the second line the graded version of the gauge field strength is shown. The gauge field strength can be expressed in the same notation as that in Eq.~(\ref{ccwz2}), i.e. distinguishing between the preserved and the broken parts:
\beq
f^+_{\mu\nu} = \frac{f_{\mu\nu}+f^{\cR}_{\mu\nu}}{2}\,, \qquad \qquad
f^-_{\mu\nu} = \frac{f_{\mu\nu}-f^{\cR}_{\mu\nu}}{2} \,.
\eeq
The preserved part of the field strength $f^+_{\mu\nu}$ and the covariant field $e_{\mu\nu}$ are related by an identity,
\beq
e_{\mu\nu} = f^+_{\mu\nu} -i\,\left[ d_\mu , d_\nu \right] \,,
\eeq
and, as a consequence, there is a certain degree of freedom in the choice of the building blocks necessary to construct the Lagrangian: two distinct sets of covariant objects can be adopted, either $\{f^+_{\mu\nu},f^-_{\mu\nu},d_\mu\}$ or $\{e_{\mu\nu},f^-_{\mu\nu},d_\mu\}$.

Since $SO(4)$ is isomorphic to $SU(2)_L\times SU(2)_R$, the custodial symmetry is embeddable in this model. However, as it is explicitly broken by the gauging of the SM group, the left and the right components of the covariant objects defined just above can be treated independently, adding more freedom in writing the effective Lagrangian. The following structures can then be introduced:
\beq
\begin{aligned}
f^+_{\mu\nu} &= f^L_{\mu\nu} + f^R_{\mu\nu}\,,\qquad \qquad 
\hat{f}^+_{\mu\nu} = f^L_{\mu\nu} - f^R_{\mu\nu} \,,\\
e_{\mu\nu} &= e^L_{\mu\nu} + e^R_{\mu\nu}\,, \qquad\qquad 
\hat{e}_{\mu\nu} = e^L_{\mu\nu} - e^R_{\mu\nu} \,,
\end{aligned}
\label{fandfhatContino}
\eeq
with the obvious relations 
\beq
e^L_{\mu\nu} = f^L_{\mu\nu} -i\,\left[ d_\mu , d_\nu \right]_L\,,\qquad\qquad
e^R_{\mu\nu} = f^R_{\mu\nu} -i\,\left[ d_\mu , d_\nu \right]_R \,.
\eeq
These covariant terms complete the list of building blocks necessary to write the effective chiral Lagrangian up to four derivatives for the $SO(5)/SO(4)$ model in the $\Omega$-representation:
\begin{itemize}
\item[i)]
The kinetic term for the GBs is described by the operator
\beq
{\mathcal L}^{(2)}=\frac{f^2}{4}\Tr\left(d_{\mu} d^{\mu}\right)\,.
\eeq
\item[ii)]
The kinetic terms for the gauge fields are described by the operator
\beq
\cO_k = \Tr[f_{\mu\nu} f^{\mu\nu}] = \Tr[f^L_{\mu\nu} f_L^{\mu\nu}] + \Tr[f^R_{\mu\nu} f_R^{\mu\nu}] + 
              \Tr[f^-_{\mu\nu} f_-^{\mu\nu}] \equiv \iL^2 + \iR^2 +\iB^2\, 
\eeq
where the definition for $\iL$, $\iR$, and $\iB$ can be easily deduced. In the following, the compact notation $\iL$, $\iR$, and $\iB$ will be adopted for shortness when necessary. 
\item[iii)]
The following two operators describing gauge-GB interactions,
\beq
{\cal O}_1 = \Tr\left(d_{\mu} d^{\mu}\right) \Tr\left(d_{\nu} d^{\nu}\right)\,,\qquad\qquad
{\cal O}_2 = \Tr\left(d_{\mu} d_{\nu}\right) \Tr\left(d^{\mu} d^{\nu}\right) \,, 
\eeq
belong to the operator basis presented both in Ref.~\cite{Contino:2011np} and in Ref.~\cite{Azatov:2013ura}. 
\item[iv-a)]
In Ref.~\cite{Contino:2011np}, focussing only on  CP-even operators with at most four derivatives, the following list has been considered: 
\beq
\begin{aligned}
{\cal O}_3 &= \Tr\left(\hat{e}_{\mu\nu} e^{\mu\nu} \right) = \Tr\left((e^L_{\mu\nu})^2-(e^R_{\mu\nu})^2 \right)\,,   \\
{\cal O}^+_4 &= i\,\Tr\left(f^+_{\mu\nu}\,[d^\mu, d^\nu]\right) 
              = i\,\Tr\left((f^L_{\mu\nu}+f^R_{\mu\nu})\,[d^\mu, d^\nu]\right)\,,\\
{\cal O}^-_4 &= i\,\Tr\left(\hat{f}^+_{\mu\nu}\,[d^\mu, d^\nu]\right) 
              = i\,\Tr\left((f^L_{\mu\nu}-f^R_{\mu\nu})\,[d^\mu, d^\nu]\right)\,, \\
{\cal O}^+_5 &= \Tr\left((f^-_{\mu\nu})^2\right)\,,\\
{\cal O}^-_5 &= \Tr\left(\hat{f}^+_{\mu\nu} f^{+\mu\nu}\right) 
              = \Tr\left((f^L_{\mu\nu})^2-(f^R_{\mu\nu})^2 \right) \,.
\end{aligned}
\label{continoOP}
\eeq
Although two additional operators with four $d_\mu$ fields could be included in general,
\beq
\begin{aligned}
{\cal O}^+_{1a} &= \Tr\left([d_\mu, d_\nu] \, [d^\mu, d^\nu]\right) =\Tr\left(([d_\mu, d_\nu]_L)^2 + ([d_\mu, d_\nu]_R)^2 \right)\,,\\
{\cal O}^-_{1a} &= \Tr\left(([d_\mu, d_\nu]_L)^2 - ([d_\mu, d_\nu]_R)^2 \right) \,,
\end{aligned}
\label{newOP}
\eeq
in the particular case of $SO(5)/SO(4)$ CH model these operators are redundant or vanishing: 
\beq
{\cal O}^+_{1a} =  \frac{1}{2} \left({\cal O}_2-{\cal O}_1\right)\,,\qquad\qquad
{\cal O}^-_{1a} = 0\,.
\eeq
It is useful to rewrite the operators in Eqs.~(\ref{continoOP}) and (\ref{newOP}) in terms of the  $SU(2)_L \times SU(2)_R$ projections. Defining for shortness, 
\beq
\begin{aligned}
{\tt L D_L} &= i\,\Tr\left(f^L_{\mu\nu}\,[d_\mu, d_\nu]_L\right)\,, \qquad \qquad
&{\tt R D_R} &= i\,\Tr\left(f^R_{\mu\nu}\,[d_\mu, d_\nu]_R\right) \\
{\tt D_L^2} &= \Tr\left([d_\mu, d_\nu]_L\,[d_\mu, d_\nu]_L\right)\,, \qquad \qquad
&{\tt D_R^2} &= \Tr\left([d_\mu, d_\nu]_R\,[d_\mu, d_\nu]_R\right) \,,
\end{aligned}
\eeq
it is possible to write:
\beq
\begin{aligned}
{\cal O}_3 &= \iL^2-\iR^2-2\,({\tt L D_L} - {\tt R D_R}) + ({\tt D_L^2} - {\tt D_R^2})\,,  \\ 
{\cal O}^+_4 &= ({\tt L D_L} + {\tt R D_R})\,,  \\
{\cal O}^-_4 &= ({\tt L D_L} - {\tt R D_R}) \,, \\
{\cal O}^+_5 &= \iB^2\,,  \\
{\cal O}^-_5 &= \iL^2 - \iR^2 \,, \\
{\cal O}^+_{1a} &= {\tt D^2_L} + {\tt D^2_R} = \frac{1}{2} \left({\cal O}_2-{\cal O}_1\right)\,,  \\
{\cal O}^-_{1a} &= {\tt D^2_L} - {\tt D^2_R} = 0 \,.
\end{aligned}
\label{ContinoBasisSO5SO4first}
\eeq
The set of operators $\{{\mathcal L}^{(2)}, \cO_k, \cO_1, \cO_2, \cO_4^+, \cO_4^-,\cO_5^+, \cO_5^-\}$ constitutes a basis for the $SO(5)/SO(4)$ CH model, while the invariants $\{\cO_3, \cO_{1a}^+, \cO_{1a}^-\}$ are redundant or vanishing. In particular, contrary to what is stated in Ref.~\cite{Contino:2011np}, $\cO_3$ is not an independent operator of the basis as it can be expressed as a linear combination of other operators:
\beq
\cO_3 = \cO^-_5 - 2\,\cO^-_4\,.
\eeq
\item[iv-b)]
The operator basis for the $SO(5)/SO(4)$ CH model presented in Ref.~\cite{Azatov:2013ura} is slightly different. Besides 
${\mathcal L}^{(2)}$, $\cO_k$, $\cO_1$ and $\cO_2$, the operators in Eq.~(\ref{continoOP}) have been substituted by the following ones\footnote{The operators of the basis in Ref.~\cite{Azatov:2013ura} will be denoted with a `` ${}'$ '' to avoid confusion with the ones in Ref.~\cite{Contino:2011np}.}:
\beq
\begin{aligned}
\cO^{\prime+}_3 &= \Tr\left(e_{\mu\nu}\,e^{\mu\nu} \right) =\Tr\left((e^L_{\mu\nu})^2+(e^R_{\mu\nu})^2 \right)\,,  \\
\cO^{\prime-}_3 &\equiv \cO_3 = \Tr\left(\hat{e}_{\mu\nu}\,e^{\mu\nu}\right)=\Tr\left((e^L_{\mu\nu})^2-(e^R_{\mu\nu})^2 \right)\,, \\
\cO^{\prime+}_4 &= i\,\Tr\left(e_{\mu\nu}\,[d^\mu, d^\nu]\right) = i\,\Tr\left((e^L_{\mu\nu}+e^R_{\mu\nu})\,[d^\mu, d^\nu]\right)\,, \\
\cO^{\prime-}_4 &= i\,\Tr\left(\hat{e}_{\mu\nu}\,[d^\mu, d^\nu]\right) = i\,\Tr\left((e^L_{\mu\nu}-e^R_{\mu\nu})\,[d^\mu, d^\nu]\right)\,, \\
\cO^{\prime}_5 &\equiv \cO^-_{1a} = \Tr\left(([d_\mu, d_\nu]_L)^2-([d_\mu, d_\nu]_R)^2\right) \,.
\end{aligned} 
\label{continoOPN}
\eeq
By rewriting these operators in terms of the  $SU(2)_L \times SU(2)_R$ projections, it follows that 
\beq
\begin{aligned}
\cO^{\prime+}_3 &= \iL^2+\iR^2-2\,({\tt L D_L} + {\tt R D_R}) + ({\tt D_L^2} + {\tt D_R^2})\,, \\
\cO^{\prime-}_3 &= \iL^2-\iR^2-2\,({\tt L D_L} - {\tt R D_R}) + ({\tt D_L^2} - {\tt D_R^2})\,, \\
\cO^{\prime+}_4 &= ({\tt L D_L} + {\tt R D_R}) + {\tt D^2_L} + {\tt D^2_R}\,, \\
\cO^{\prime-}_4 &= ({\tt L D_L} - {\tt R D_R}) + {\tt D^2_L} - {\tt D^2_R}\,, \\
\cO^{\prime}_5 &= {\tt D^2_L} - {\tt D^2_R}=0\,.
\end{aligned}
\eeq
With respect to the basis in Ref.~\cite{Contino:2011np}, the operator $\cO^{\prime-}_3 = \cO_3$ should now be taken as part of the basis, as it is the only one containing the combination $L^2-R^2$ ($\cO^-_5$ does not have a counterpart in this basis). 
The total number of operators entering the basis is the same as in the previous case: $\{{\mathcal L}^{(2)}, \cO_k, \cO_1, \cO_2, \cO^{\prime +}_3, \cO^{\prime-}_3,\cO^{\prime+}_4, \cO^{\prime-}_4\}$, as $\cO^{\prime}_5$ is automatically vanishing.
\end{itemize}

\subsection{Comparison with the basis in Ref.~[49]}

As described in Ref.~\cite{Contino:2011np}, the EWSB is induced due to a misalignment between the $SO(4)$ subgroup, left unbroken in the global $SO(5)$ breaking, and the $SO(4)'$ subgroup that contains the SM gauged group. A rotation between these two directions can be defined and it can be parametrised by an angle $\theta$. Accordingly, the $SO(4)$ generators and the $SO(4)'$ ones are connected to each other through the rotation $R_\theta$ and, to recover the results in Eq.~(13) of Ref.~\cite{Contino:2011np}, the following relation between the GB matrices $U$, introduced in appendix \ref{App:Contino},  and $\Omega$ should be adopted:
\beq
U = \Omega\,R_\theta^\dag\,,
\eeq
and it follows that
\beq
U^\dag \DL_\mu U = R_\theta\left(\Omega^\dag \DL_\mu \Omega\right) R^\dag_\theta
\quad \longrightarrow \quad
d_\mu = -\frac{i}{2}R_\theta\, v_\mu\, R^\dag_\theta\,, \quad
e_\mu = R_\theta\, p_\mu\, R^\dag_\theta\,.
\eeq
Furthermore, a link between the gauge field strengths $s_{\mu\nu}$ and $f_{\mu\nu}$ can be found:
\beq
f_{\mu\nu} = R_\theta\, s_{\mu\nu}\, R^\dag_\theta\,.
\eeq
It is now possible to identify the relation among the operator basis in Ref.~\cite{Contino:2011np} and reported in Eq.~(\ref{ContinoBasisSO5SO4first}) (excluding the redundant operator $\cO_3$) and the operators in Eqs.~(\ref{BasisCPinOmega}) and (\ref{BasisCBinOmega}). A similar discussion can be performed for the operators basis in Ref.~\cite{Azatov:2013ura}. Focussing to the case in which the full group $SO(5)$ is gauged, i.e. any source of custodial breaking, SM or beyond, is neglected, it follows that the two bases are equivalent:
\beq
\begin{aligned}
{\mathcal L}^{(2)}&\rightarrow -\frac{f^2}{16}\Tr\left(v_\mu\,v^\mu\right)\,,\\
\cO_k&\rightarrow \Tr\left(s_{\mu \nu}s^{\mu \nu}\right)\,,\\
\cO_1&\rightarrow \frac{1}{16}\Tr\left(v_\mu\,v^\mu\right)\Tr\left(v_\mu\,v^\mu\right)\,,\\
\cO_2&\rightarrow \frac{1}{16}\Tr\left(v_\mu\,v_\nu\right)\Tr\left(v^\mu\,v^\nu\right)\,,\\
\cO_4^+& \rightarrow -\frac{i}{4}\Tr\left(s_{\mu\nu}\left[v^\mu,v^\nu\right]\right)\,,\\
\cO_5^+& \rightarrow 2\Tr\left(s_{\mu \nu}s^{\mu \nu}\right)-2\Tr\left(s^\cR_{\mu\nu}s^{\mu\nu}\right)\,.
\end{aligned}
\eeq
Indeed the last three operators in Eq.~(\ref{BasisCPinOmega}), that do not appear in the list above, are vanishing or redundant due to the fact that fermions are neglected in Ref.~\cite{Contino:2011np} and due to the algebra of the $SO(4)$ generators (see appendix~\ref{App:Contino}).

In the case in which only the SM symmetry is gauged, introducing then an explicit breaking of the custodial symmetry due to the hypercharge group, the two bases do not coincide anymore:
\beq
\begin{aligned}
{\mathcal L}^{(2)}&\rightarrow \frac{1}{4}\cAt_C\,,\\
\cO_k&\rightarrow \cAt_B+\cAt_W\,,\\
\cO_1& \rightarrow \frac{1}{16}\cAt_4\,,\\
\cO_2& \rightarrow \frac{1}{16}\cAt_5\,,\\
\cO_4^+& \rightarrow -\frac{1}{4}\left(\cAt_2+\cAt_3\right)\,,\\
\cO_4^-& \rightarrow \frac{1}{4}\left(\cAt_2-\cAt_3\right)\,,\\
\cO_5^+& \rightarrow 2\left(\cAt_B+\cAt_W\right)-2\left(\cAt_{B\Sigma}+\cAt_{W\Sigma}+2\cAt_1\right)\,,\\
\cO_5^-&\rightarrow -\cAt_B+\cAt_W\,.
\end{aligned}
\label{LinkAmongUsAndContinoFirst}
\eeq
The operators ${\mathcal L}^{(2)}$, $\cO_1$ and $\cO_2$ are in a one-to-one correspondence with the operators $\cAt_C$, $\cAt_4$ and $\cAt_5$; the two operators $\cO_k$ and $\cO_5^-$ ($\cO_4^+$ and $\cO_4^-$) are connected to two linear independent combinations of $\cAt_B$ and $\cAt_W$ ($\cAt_2$ and $\cAt_3$); finally, the operator $\cO_5^+$ is connected with a linear combination of five operators of the basis in Eq.~(\ref{BasisCBinOmega}), identifying therefore a relation among $\cAt_{B\Sigma}$, $\cAt_{W\Sigma}$ and $\cAt_1$. 

In summary, the analysis in this appendix has clarified the connection with previous literature. The differences between the basis presented here and that in Ref.~\cite{Contino:2011np} are understood in terms of the different sources of custodial breaking assumed:
\begin{itemize}
\item[-]  Ref.~\cite{Contino:2011np} describes an explicit breaking of the $SO(4)$ subgroup (this is encoded in their definition of $\hat{f}^+_{\mu\nu}$). As a result, their basis is composed of eight independent operators.
\item[-] This work instead implements an explicit breaking of $SO(4)'$ in the language of Ref.~\cite{Contino:2011np}. This originates  from treating independently the gauge fields $\WL_\mu$ and $\BL_\mu$. This choice closely follows the approach of Appelquist and Longhitano of considering all possible SM $SU(2)_L\times U(1)_Y$ invariant operators:  operators $\cAt_{W\Sigma}$ and $\cAt_{B\Sigma}$ arise then as independent structures. As a consequence, the basis requires ten independent operators.
\end{itemize}


\providecommand{\href}[2]{#2}\begingroup\raggedright\endgroup

\end{document}